\documentclass[prd,aps,floatfix,nofootinbib,preprint ,tightenlines,superscriptaddress]{revtex4}
\usepackage{dcolumn}
\usepackage{amsmath}
\usepackage{latexsym}
\usepackage{graphicx}
\usepackage{bm}

\def\co{{\cal O}}
\def\bx{{\bf x}}

\def\svev#1{\left\langle #1\right\rangle}       

\def\Tr{{\rm Tr}\,}

\newcommand{\bee}{\begin{equation}}
\newcommand{\ee}{\end{equation}}
\newcommand{\beea}{\begin{eqnarray}}
\newcommand{\eea}{\end{eqnarray}}

\begin{document}
\title{Finite temperature properties of QCD with two flavors and three, four, and five colors}

\author{Thomas DeGrand}
\affiliation{
Department of Physics, University of Colorado,
        Boulder, CO 80309 USA}
\email{thomas.degrand@colorado.edu}

\date{\today}

\begin{abstract}
I present a numerical study  of the crossover  between the low temperature chirally broken phase and the high temperature chirally restored phase
 in $SU(N_c)$
gauge theory with $N_c=3-5$ colors and $N_f=2$ degenerate fermion flavors.
Fermion masses span a range of intermediate values
corresponding to pseudoscalar to vector meson masses
$(m_{PS}/m_V)^2\sim 0.25$ to 0.63.
 Observables include the temperature dependent chiral condensate and screening masses.
At each fermion mass these quantities  show nearly identical
temperature dependence across $N_c$.
\end{abstract}
\maketitle

\section{Introduction and motivation  \label{sec:intro}}
 
The limit of QCD when the number of colors $N_c$ is taken large
has a long history as a source of insight about $N_c=3$ QCD itself  \cite{tHooft:1973alw,tHooft:1974pnl,Witten:1979kh} .
Many of the applications of large
 $N_c$ ideas to phenomenology
actually involve nonperturbative quantities (masses or decay constants), although the predictions are
often based on counting the color weight of Feynman diagrams.
To what extent are these predictions true?
Checking them requires a lattice simulation, and there is a small literature of lattice calculations
away from $N_c=3$ to provide such tests.
(See  Refs.~\cite{Lucini:2012gg,GarciaPerez:2020gnf,Hernandez:2020tbc}
for a selection of reviews.)

At a qualitative level, lattice calculations confirm large $N_c$ intuition
rather nicely: meson masses, baryon masses, and decay constants scale as $N_c^p f(m_q)$
where $p$ is a characteristic power and $m_q$ is the fermion mass.
The interplay of small dynamical fermion mass and large $N_c$ is less well explored,
but simple matrix elements (decay constants,  the kaon weak matrix element
calculations of Refs~\cite{Donini:2016lwz,Donini:2020qfu}) mostly scale as expected. So do some chiral 
observables which are governed by the pseudoscalar decay constant
(scaling as   $m_{PS}^2/f_{PS}^2 \propto m_{PS}^2/N_c$:  compare to Ref.~\cite{Hernandez:2019qed}) or by the condensate
(as in the topological susceptibility $\chi_T \propto m_q \Sigma \propto m_q N_c$ \cite{DeGrand:2020utq}).

The subject of this paper is the crossover temperature for the transition from the low temperature
confining and chirally broken phase to the high temperature deconfined and chirally restored phase,
for QCD with $N_c=3$, 4, and 5 colors and $N_f=2$ flavors of degenerate mass fermions in the fundamental
representation. The reason why this project might be interesting is that there
is not a  single large $N_c$ story for what happens, there are at least three possibilities.

The first possibility comes from the naive large $N_c$ expectation that gluonic degrees of freedom dominate
fermionic ones as $N_c \rightarrow \infty$. There should be a finite temperature
confinement  deconfinement transition which converges to the pure glue  one in the
large $N_c$ limit. This is a first order transition with $T_c \sim 320$ MeV.
In $N_c=3$ the pure gauge transition is first order and as the fermion
mass falls from infinity the transition becomes a crossover. In this scenario the large $N_c$
transition  temperature  would remain roughly constant across $N_c$ as the fermion mass fell from infinity,
and the critical point where the first order region ends would move to smaller fermion mass as $N_c$ rises.

The second scenario assumes naive chiral symmetry breaking dominance. Even QCD at large $N_c$
has an $SU(N_f) \times SU(N_f)$ symmetry which (modulo issues with the eta-prime \cite{Kaiser:2000gs})
undergoes spontaneous symmetry breaking to vectorial $SU(N_f)$. The Pisarski - Wilczek analysis
 \cite{Pisarski:1983ms} approximates the Goldstone sector
as a linear sigma model. For $N_f=2$ the system is expected to have a second order transition
at zero fermion mass, with $O(4)$ critical exponents. Second order transitions are unstable under
perturbation, so the transition becomes a crossover away from $m_q=0$.
All QCDs with any $N_c$   should then share
a common behavior at small fermion mass.

This paper doesn't have data at the tiny (or zero) fermion masses needed to say anything about
the properties of any transition at zero fermion mass, but it can ask a question which is
related to the Pisarski - Wilczek analysis: how does the crossover temperature scale with $N_c$?
Linear sigma models contain one dimensionful parameter, the vacuum expectation value of the
scalar field, and all derived  dimensionful quantities (the pseudoscalar decay constant, and the
crossover temperature $T_c$ itself) are proportional to it. It is well known from previous
large $N_c$ spectroscopy comparisons that $f_{PS} \propto\sqrt{N_c}$. Thus the naive prediction
of the second scenario is $T_c \propto \sqrt{N_c}$ \cite{Pisarski}.  Is it so?

The final scenario predates QCD. Confining theories are expected to show an exponentially growing spectrum of
resonances with mass, forming a Hagedorn spectrum  \cite{Hagedorn:1968zz}.
  The tower of resonances implies a limiting temperature
$T_0$ and (as first stated explicitly by Cabibbo and Parisi \cite{Cabibbo:1975ig}, as far as I  can tell) this implies
a crossover temperature $T_c \sim T_0$. In QCD, the Hagedorn temperature is about 160 MeV.
The extension of the story for large $N_c$ and nonzero $N_f$ is that the spectrum of resonances
is basically identical across $N_c$. Meson states dominate baryon ones up to a few GeV. If two 
theories have the same spectrum, then they will
have the same critical properties.  Notice that large $N_c$ with nonzero $N_f$ is different from quenched QCD:
the latter case has a much sparser spectrum below 2 GeV. There 
are only glueballs  in contrast to all the excited states labeled (for example) by quark model counting.
The prediction of the third scenario is that  any $N_c \ne 3$
with $N_f=2$ will qualitatively resemble $N_c=3$, $N_f=2$. This argument  has been used to
justify the observation that the deconfinement transition in pure gauge $SU(N_c)$ is nearly
independent of $N_c$ \cite{Lucini:2005vg,Lucini:2012wq}, since the glueball spectrum is also nearly $N_c$ independent.

The first scenario was already unlikely given that the deconfinement critical point for $N_c=3$ is already at a very high mass.
Ref~\cite{Cuteri:2020yke}
observes it at a pseudoscalar mass of about 4 GeV. And indeed, there is no evidence  in any of the simulations
reported here for anything other than a smooth crossover. (This 
makes the identification of a particular crossover temperature problematic.)
As to the other scenarios, my results indicate that the temperature dependence of observables computed at common values of the
fermion mass show essentially identical behavior, including inflection points at finite temperature, which is
 nearly independent of $N_c$.

\begin{figure}
\begin{center}
\includegraphics[width=0.9\textwidth,clip]{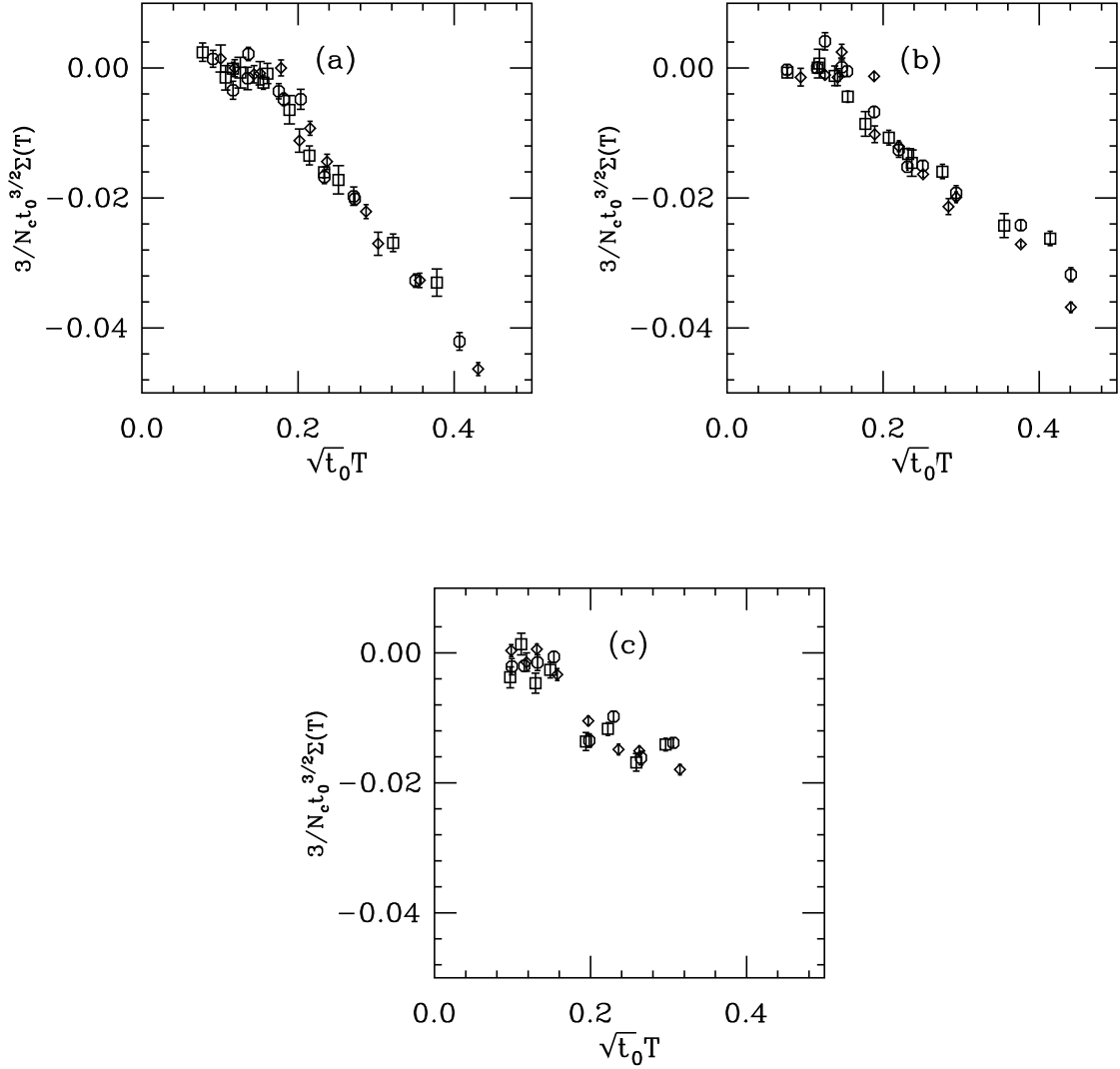}
\end{center}
\caption{The temperature dependent condensate, rescaled by $3/N_c$,
as a function of temperature, in appropriate units of $t_0$.
Squares, octagons, and diamonds label $N_c=3$, 4, and 5.
(a)  $(m_{PS}/m_V)^2 \sim 0.63$; 
(b) $(m_{PS}/m_V)^2 \sim 0.5$;
(c) $(m_{PS}/m_V)^2 \sim 0.25$.
\label{fig:sigmaall}}
\end{figure}

To illustrate this  statement, the
 temperature dependent and zero  temperature subtracted condensate, defined below in Eq.~\ref{eq:mycond},
is shown in Fig.~\ref{fig:sigmaall}. The naive $N_c$ scaling of the condensate is divided out, and 
data is presented for three values of the ratio $(m_{PS}/m_V)^2$. This quantity is close to zero at low temperature
and becomes negative at high temperature as (speaking loosely) chiral symmetry is restored and the finite temperature
condensate falls to zero. The different plotting symbols label different numbers of colors.
This picture illustrates the smooth crossover from broken to restored chiral symmetry with nearly identical temperature
variation across $N_c$. 

A natural question to ask is, can one identify a feature in the data which serves as a marker for a crossover temperature?
The short answer is no. The quantities I have studied have a smooth temperature dependence
with no sharp features.  I was able to identify two quantities
which could serve as markers: there is a peak in the derivative $d\Sigma(T)/dT$, and there is a transition region
where screening masses  cross over from near independence with temperature to
strong temperature dependence. Both features are broad.

 The outline of the rest of the paper is as follows:
 Technical aspects of the calculation (lattice action, data sets, a bit about data analysis methodology) are described in Sec.~\ref{sec:tech}.
 This is all completely conventional. Then the various observables are described and results are presented for them:
 the temperature dependent condensate in Sec.~\ref{sec:tcondens},
 screening masses in Sec.~\ref{sec:screen}. I mention the
 the Polyakov line susceptibility in Sec.~\ref{sec:poly}.
 Conclusions are summarized in Sec.~\ref{sec:conclusions}.

\section{Technical aspects of the calculation \label{sec:tech}}

\subsection{Simulation methodology,  lattice actions, data sets \label{sec:action}}

The simulations involved two degenerate flavors of
Wilson--clover fermions. The gauge action is the usual Wilson plaquette action.
The fermion action uses gauge connections defined as normalized hypercubic 
 smeared links~\cite{Hasenfratz:2001hp,Hasenfratz:2007rf,DeGrand:2012qa,DeGrand:2016pur}.
The bare gauge coupling $g_0$ is set by the simulation parameter $\beta = 2N_c / g_0^2$.
The bare quark mass $m_0^q$ is introduced via the hopping
parameter $\kappa=(2m_0^q a+8)^{-1}$. The clover coefficient is fixed to
 its tree level value, $c_{\text{SW}}=1$.
Gauge-field updates used the Hybrid Monte Carlo (HMC)  algorithm \cite{Duane:1986iw,Duane:1985hz,Gottlieb:1987mq}
with a multi-level Omelyan integrator \cite{Takaishi:2005tz} and
multiple integration time steps \cite{Urbach:2005ji}
with one level of mass preconditioning for the fermions \cite{Hasenbusch:2001ne}.
Lattices used for analysis are spaced a minimum of 10 HMC time units apart.
All data sets  are based on a single stream for each set of bare parameters.

The only new feature to report in these simulations is a first order bulk transition in strong coupling for $SU(5)$.
Its location depends, of course, on the particular form of the bare action.
These transitions are well known features of the pure gauge systems and the one I found
has a small $\kappa$ limit which
appears to approach the value of gauge coupling where Ref.~\cite{Lucini:2005vg}
observed a pure gauge transition.
I only mention this in passing;  data sets used to do physics were selected to avoid this
transition.

The choice of Wilson-clover fermions in a finite temperature study is not optimal,
since the best signals for finite temperature behavior with dynamical fermions are ones which are sensitive to chiral symmetry.
The reason I chose this discretization is that I already had a large collection of zero temperature
simulations which could be used to find lines of constant physics in bare parameter space,
and because I was only looking for gross finite temperature features.

\subsection{Fixing the lattice spacing \label{sec:flow}}

The lattice spacing is set by the Wilson flow parameter $t_0$ \cite{other,Luscher:2010iy}, and quantities will 
be presented as dimensionless ones by a rescaling by an appropriate power of $t_0$.

The determination of $t_0$ is done on  
zero temperature lattices, in the standard way, from the observable
$E(t_0)$  extracted from the field strength tensor,
\bee
t_0^2 \svev{E(t_0)} = C(N_c)  .
\label{eq:flow}
\ee
 $C(N_c)$ is chosen to match what most other large $N_c$
simulations take,
\bee
 C(N_c)= C \left( \frac{3}{8} \frac{N_c^2-1}{N_c}\right),
\label{eq:ce}
\ee
with $C=0.3$  the usual value used in  $SU(3)$.
The extraction of $t_0$ from the data is identical to what was done in Ref.~\cite{DeGrand:2020utq}
and details may be found there.

I use values of $t_0$ computed at each bare parameter value, that is, there is no extrapolation  or interpolation in  fermion mass.
Let's  recall a few useful numbers which will place results in context.
 The critical temperature for pure gauge systems was published in  Refs.~{\cite{Lucini:2005vg,Lucini:2012wq}}.
  The authors of these papers quote $T_c/\sqrt{\sigma}$ where $\sigma$ is the string tension.
 I converted their numbers to $\sqrt{t_0} T_c$ using  the Sommer parameter $r_0 \sim 0.49$ fm
 \cite{Sommer:1993ce},  $r_0 \sqrt{\sigma}=1.175$ for quenched $SU(3)$ and $SU(5)$ from  Ref.~\cite{DeGrand:2012hd}, 
 and the quenched $N_c=3$   $\sqrt{t_0}=0.1638$ fm from  Refs.~\cite{Lottini:2013rfa,Bruno:2013gha}  
 (as quoted in Ref.~\cite{Sommer:2014mea}) to give the pure gauge transition at 
 $\sqrt{t_0}T_c=0.265$, 0.261 and 0.278 for $N_c=3$, 4, 5.
 
 The nominal  ``physical point'' value (with $2+1$ flavors) of fermions with   $T_c=150$ MeV, $\sqrt{t_0}= 0.147$ fm
 of Ref.~{\cite{Soltz:2015ula}} from chiral observables, is $\sqrt{t_0}T_c=0.11$.

\subsection{Data sets \label{sec:sets}}

At each $N_c$ the bare parameter space is (at least) three dimensional, involving a bare gauge coupling, a bare fermion mass expressed
through $\kappa$  in units of the lattice spacing as $am_q$, and the temperature, $T=1/(aN_t)$ where $N_t$ is the length of the lattice
in the Euclidean temporal direction. 
The issue for a study like this is that no single simulation (at any one set of bare parameters) is interesting in itself: 
all features are broad. It is hard to avoid generating data sets at many parameters and easy to get lost
wading through them.

The literature suggests at least two ways to proceed.

The first approach is to map out a crossover line (or phase boundary, if it exists) as a function of bare parameters at fixed $N_t$ values.
Then zero temperature simulations are done along the crossover lines to determine physical quantities such as
the lattice spacing and $(m_{PS}/m_V)^2$.
(Examples of this approach I found useful were Refs.~\cite{AliKhan:2001ek,Ejiri:2009hq,Bornyakov:2009qh}.)
The issue with this approach is that since there is no true phase transition, the crossover region is broad, and
to present a result such as $\sqrt{t_0}T_c$ versus $(m_{PS}/m_V)^2$ involves collecting a
large number 
of zero temperature data sets followed by interpolation of their output onto a surface in bare coupling constant space.
The surface's  own bare parameters are poorly defined because the ``transition'' is just a crossover.
This issue is compounded of course by the need to work at several values of $N_c$.
 The preliminary version of this project, Ref.~\cite{DeGrand:2018tzn} used this approach, but I found it to be unwieldy.
 
 Instead, I took an approach inspired by
 Refs.~\cite{Borsanyi:2012uq,Borsanyi:2015waa}. I started with zero temperature data at selected bare parameters
 and then varied the temperature by varying $N_t$ holding all other bare parameters (and hence, the lattice spacing, $t_0$ and
 $(m_{PS}/m_V)^2$) fixed.
 A disadvantage of this approach is that it is unlikely that one will obtain data precisely at a crossover temperature.
 This can be compensated for, to some extent, by combining data sets from several values of the lattice spacing
 (but the same $(m_{PS}/m_V)^2$ values) to fill in the curves. This of course introduces another disadvantage: the data sets from different values
 of the lattice spacing have different lattice artifacts. The naive approach of just plotting them all together ignores lattice artifacts: they
 will appear in the scatter between points on a graph.
 An advantage of the approach is that it allows one to use a ``natural'' observable -- the temperature -dependent
 condensate -- to compare crossover behavior across $N_c$. 
 At the end, I have eight bare parameter sets per $N_c$
 [divided into three different $(m_{PS}/m_V)^2$ values],  five values of $N_t$ per bare parameter,
 -- 40 sets per $N_c$, 120 datasets in all.

 For zero temperature data sets I took $16^3\times 32$
volumes. Results for many of the parameter sets have been published before (see Refs.\cite{DeGrand:2016pur,DeGrand:2020utq})
and additional sets were generated to give several lattice spacings at three matching points in the fermion mass.
The squared ratio of the pseudoscalar to vector  masses  $(m_{PS}/m_V)^2$  will be taken as a proxy for a fermion mass.
and data were collected at
  $r=(m_{PS}/m_V)^2\sim 0.63$,
0.5 and about 0.25 (in the latter case, ranging from 0.22-0.27).
Useful results from the zero temperature data sets are summarized in Table \ref{tab:datasets}.

The calculation of a temperature dependent condensate described below in Sec.~\ref{sec:tcondens}
requires data sets at fixed spatial volume (for each lattice spacing), again  $16^3$ sites.
I varied $N_t$ from 4 to 12 at $r=0.63$ and $r=0.5$, and 6 to 16 at $r=0.25$ since the crossover temperature
appeared to fall with decreasing $r$. Almost all of these finite temperature data sets are 1600 trajectories long after equilibration,
again saving lattices spaced 10 trajectories apart for further analysis.
The results for screening masses and related quantities are based on subsets of these data sets since these results were not used
for any precise calculations.

\subsection{Data analysis \label{sec:autocorr}}

Most results are global observables, a single quantity averaged over the simulation run.
 For most observables, there are not really enough data for
a reliable determination of an integrated autocorrelation time.
Instead, autocorrelation times are estimated and errors are assigned  from a jackknife analysis 
dropping successive measurements from the data stream.
 For an observable $Q$ I compute  an average $\svev{Q}$ and a susceptibility 
  $\chi_Q =\svev{Q^2}-\svev{Q}^2$; the uncertainty of each comes from a jackknife.
  I varied the size of the jackknife (dropping $n_J$ successive measurements).
Typically $\Delta Q$ rises as $n_J$ rises, reflecting the effect of time autocorrelations,
until it plateaus as $n_J$ approaches or exceeds the autocorrelation time.
 At the same time  the uncertainty in $\Delta Q$
 has a fractional error from loss of statistics,
\bee
\frac{\Delta (\Delta Q)}{\Delta Q} = \sqrt{\frac{2}{n}}
\label{eq:deltaq}
\ee
where $n= N/n_J$ for $N$ measurements. The increasing uncertainty in $\svev{\Delta Q}$
eventually exceeds the variation of $\Delta Q$ with $n_J$.
The uncertainties quoted in the paper come from  jackknife averages taking $n_J$ where the quantity
$\Delta (\Delta Q)$ from Eq.~\ref{eq:deltaq} is larger than the apparent rise in the jackknife estimated $\Delta Q$.

Generally, I observed that for observables related to the temperature dependent condensate
 the errors taken from  jackknifes dropping one or two
successive lattices (spaced 10 molecular dynamics time steps apart) were indistinguishable within
the uncertainty of Eq.~\ref{eq:deltaq}.
Volume  averaged Polyakov lines were measured every trajectory in the normal course of data collection.
There, Eq.~\ref{eq:deltaq} indicated that typically jackknife errors saturated  with $n_J=5-20$ (the higher
number coming closer to the crossover temperature). The errors quotes in tables and shown in graphs
were taken from this procedure.

\section{The temperature dependent condensate and related quantities\label{sec:tcondens}}

\subsection{Defining the condensate\label{sec:tcondens1}}

With Wilson fermions the chiral condensate is a bit awkward to measure \cite{Borsanyi:2012uq,Giusti:1998wy}.
The bare condensate for Wilson fermions  at bare mass $m_1$ has an expansion
\bee
\svev{\bar \psi_0 \psi_0} = c_0 + c_1(m_1-m_0) + c_2 (m_1-m_0)^2+\dots
\ee
where $m_0$ is the bare mass at which  the axial Ward identity
fermion mass (defined below) vanishes, $m_1$ is the bare mass at the simulation point,
and the $c_i$'s contain cutoff (divergent) behavior, $c_0 \sim 1/a^3$, $c_1 \sim 1/a^2$, $c_2 \sim 1/a$.
The divergent pieces are independent of temperature, and so the difference
\bee
\svev{\bar \psi \psi}_{sub} = \svev{\bar \psi \psi}_T - \svev{\bar \psi \psi}_{T=0} 
\ee
is finite and sensible.
Inspired by the Gell-Mann, Oakes, Renner relation, a potential definition for $\svev{\bar \psi \psi}_{sub} $
is
\bee
\svev{\bar \psi \psi}_{sub}  \propto m_{AWI}\left[ \int d^4 x \svev{0|P(x,t) P(0,0)|0}_T - \int d^4 x \svev{0|P(x,t) P(0,0)|0}_{T=0} \right]
\label{eq:cond}
\ee
where $P(x,t) = \bar \psi(x,t) \gamma_5 \psi(x,t)$ is the pseudoscalar current. The first term on the right hand side
of Eq.~\ref{eq:cond} is evaluated on an $N_s^3\times N_t$ lattice where $T=1/(aN_T)$ and the second term is evaluated
on an $N_s^3\times N_t$ lattice where $N_t \gg N_s$.

 With the ``130 MeV'' definition of $f_{PS}$ the Gell-Mann, Oakes, Renner relation between the condensate $\Sigma$ and other observables is
\bee
\Sigma = \frac{m_{PS}^2 f_{PS}^2}{4m_q}.
\label{eq:gmor}
\ee
In this convention, with the
axial current $A_\mu^a=\bar \psi \gamma_\mu\gamma_5 (\tau^a/2)\psi$, and the pseudoscalar
density $P^a=\bar \psi \gamma_5 (\tau^a/2)\psi$,  the vacuum to pseudoscalar matrix elements are
$\svev{0|A_0| PS} = m_{PS} f_{PS}$ and  $\svev{0|P|PS} = m_{PS}^2 f_{PS}/(2m_q)$ .
The partial conservation of axial current  relation is
\bee
\partial_\mu A_\mu(x,t) = 2m_q P(x,t).
\ee
Matrix elements of this relation define $m_q$ to be the Axial Ward Identity (AWI) fermion mass, through the two-point functions
\bee
\partial_t \sum_\bx \svev{A_0^a(\bx,t)\co^a} = 2m_q \sum_\bx \svev{ P^a(\bx,t)\co^a},
\label{eq:AWI}
\ee
where  $\co^a$ can be any convenient source. 

The massive correlator in finite volume with periodic temporal boundary conditions
 in temporal length $N_t$, saturated by a single state of mass $m_{PS}$, is
\bee
C(t)= \sum_x \svev{P(x,t) P(0,0)} = |\svev{0|P|PS}|^2 \frac{\cosh (m_{PS}(N_t/2-t))}{2m_{PS}\sinh (m_{PS}N_t/2)} .
\ee
Thus its integral over the simulation volume is
\bee
\int_0^{N_t} dt C(t)= \frac{ |\svev{0|P|PS}|^2 }{m_{PS}^2} = \left(\frac{m_{PS}^2 f_{PS}}{2m_q}\right)^2 \frac{1}{m_{PS}^2} =  \frac{\Sigma}{m_q} .
\label{eq:condsub}
\ee
In the passage from lattice to continuum regularization there is a factor of $Z_A^2$  on the right hand side of Eq.
\ref{eq:condsub} where $Z_A= (1-(3\kappa)/(4\kappa_c))z_A$ is the tadpole improved $Z-$ factor; $z_A= 1+ c \alpha/(4\pi) \sim 1$.
$z_A$ is close to unity for the lattice action used here and so we omit it from further discussion.

Borsanyi et al \cite{Borsanyi:2012uq,Borsanyi:2015waa} 
wrote in the days before the use of $t_0$ and so they presented plots of
$(m_q \svev{\bar \psi \psi}_{sub})/m_{PS}^4$ as a dimensionless observable.  I instead will look at
the quantity
\bee
\frac{3}{N_c} t_0^{3/2} \Sigma(T) = \frac{3}{N_c} t_0^{3/2} \times m_q (\Delta_{PP}(T) - \Delta_{PP}(T=0) )
\label{eq:mycond}
\ee
where (explicitly showing the conversion from the lattice quantity computed with clover fermions to a continuum one)
\bee
\Delta_{PP}(T) = \hat \Delta_{PP}(N_t) (1-\frac{3\kappa}{4\kappa_c})^2 .
\ee
and the lattice quantity measured with the usual convention for the definition of lattice field variables is
\bee
\hat \Delta_{PP}(N_t) = \sum_{t=0}^{N_t} \sum_x \svev{P(x,t) P(0,0)}   .
\label{eq:delta}
\ee

The factor of $t_0^{3/2}$ in Eq.~\ref{eq:mycond}
renders the observable dimensionless and the overall factor of $3/N_c$ is included so that plots can show 
collapse to a common curve across
$N_c$ when the condensate scales proportional to $N_c$ as expected by large $N_c$ counting.

Data for the integrated pseudoscalar correlator is recorded in Tables \ref{tab:sig63}, \ref{tab:sig5} and \ref{tab:sig25}.
Figure \ref{fig:sigmaall} shows the rescaled  temperature dependent condensate from 
Eq.~\ref{eq:mycond} at three values of $r=(m_{PS}/m_V)^2=0.63$, 0.5 and about 0.25.
These plots are sufficient to show that the finite temperature behavior of the systems is
reasonably independent of $N_c$. However, the curves are too featureless
 to identify an inflection point as a signal for a crossover temperature.

 
 \subsection{Checking for effects of finite volume and nonzero lattice spacing \label{sec:checks}}

 Two potential issues with the calculation should be discussed before proceeding: the first is whether the simulation volume
 could affect the results. The second is a check of the lattice spacing dependence of the data presented in  Fig.~\ref{fig:sigmaall}. 
 
 Finite volume effects were studied in an earlier paper, Ref.~\cite{DeGrand:2016pur},
  involving some of the data sets used here. Here is a recapitulation of that analysis,
 which basically follows the treatment of Sharpe  \cite{Sharpe:1992ft}.
  Simulation volume effects typically arise from tadpole contributions due to pseudoscalar meson emission and absorption,
 where the meson returns not to its original emission point but to an image point.
The pseudoscalar correlator for a particle of mass $m$
 in a box of length $L_\mu$ in direction $\mu$ can be written as
\bee
\Delta(m,x) \rightarrow \sum_{n_\mu} \Delta(m,x+n_\mu L_\mu) .
\label{eq:fs}
\ee
The infinite volume propagator, call it $\bar \Delta(m,x)$, is the $n=0$ term in the sum.
The finite volume tadpole is
\bee
\Delta(m,0)= \bar\Delta(m,0) + \bar I_1(m,L)
\ee
where $\bar I_1(m,L)$ is the sum over images.
If a typical infinite volume observable has a chiral expansion
\bee
O(L=\infty)=O_0[1+ C_0 \frac{1}{f_{PS}^2} \bar \Delta(m,0) ]
\label{eq:typical}
\ee
then the finite volume correction is
\bee
O(L)-O(L=\infty)=O_0[ C_0 \frac{1}{f_{PS}^2} \bar I_1(m,L) ].
\label{eq:fvcorr}
\ee
Sharpe  \cite{Sharpe:1992ft}
has shown that nearest image contribution gives a useful lower bound on the finite volume correction.
It is
\bee
I_1(m,L) \sim 6\left(\frac{m^2}{16\pi^2}\right) \left( \frac{8\pi}{(mL)^3}\right)^{1/2} \exp(-mL).
\label{eq:guesstimate}
\ee
The factor of 6 counts the closest neighboring points at positive and negative offsets.

\begin{figure}
\begin{center}
\includegraphics[width=0.5\textwidth,clip]{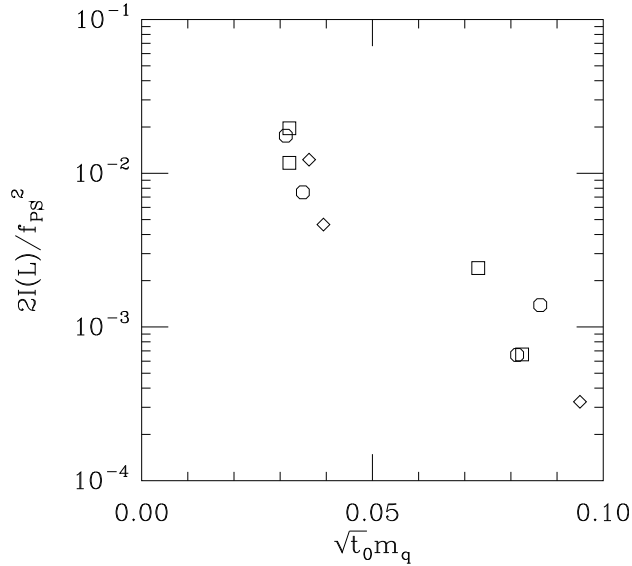}
\end{center}
\caption{Expected finite size effect from Eq.~\protect{\ref{eq:guesstimate}}, from our tabulated
data. Symbols are 
squares for $SU(3)$, octagons for $SU(4)$, diamonds for $SU(5)$. Only the small fermion mass part of the data collection is shown.
\label{fig:finite}}
\end{figure}

We can use Eq.~\ref{eq:guesstimate}, plus our tables of lattice masses and decay constants,
 to check to see which of our
data sets might be compromised by volume. The result, $2I_1(m,L)/f_{PS}^2$
(the 2 is needed to convert our 130 MeV definition of the decay constant to
the standard chiral literature's 93 MeV)  is shown in Fig.~\ref{fig:finite} for the low mass end of our data set, mostly the
sets labelled $(m_{PS}/m_V)^2\sim 0.25$.

Typical values of $C_0$ for chiral observables in Eq.~\ref{eq:typical} are order unity numbers and so this indicates that finite volume does not seem
to be an issue with this data set as long as one is not looking for precision at the  few per cent level.
Note that the usual shorthand taking $m_{PS}L$  to be greater than some minimum value is not really applicable to $N_c>3$ because of
the $1/f_{PS}^2$ in Eq.~\ref{eq:fvcorr}.

Cutoff dependence can be shown by breaking out the data in Fig.~\ref{fig:sigmaall} (and related quantities)
into plots where the $x$ axis is a measure of the lattice spacing. I will take this measure to be $a^2/t_0$,
the inverse of the lattice flow parameter. An issue with such tests is that the bare parameters are not carefully matched. A better way to proceed will
be to do a combined fit of observables related to chiral symmetry breaking to a version of chiral perturbation theory which includes lattice artifacts.
 To do this well requires many more zero temperature data sets than the ones used here, and is a topic for future work.
 
 A first indirect test is to look at the zero temperature condensate. This is useful for a check related to Fig.~\ref{fig:sigmaall}:
 One expectation would be that the curves in Fig.~\ref{fig:sigmaall} would show a sigmoid behavior
with $\Sigma(T)$ zero at low temperature since (loosely speaking) the condensate is unchanged as the temperature rises, then a fall as the
finite temperature condensate goes to zero, followed by a plateau at a constant value, basically the negative of
the zero temperature condensate. One could identify a crossover temperature as the midpoint on the sigmoid.

The issue with doing this is that $T=1/(aN_t)$ so going to high temperature
at fixed $a$ means going to smaller $N_t$, and at some point $N_t$ is so small that lattice artifacts must appear.
An alternative is to do a direct measurement of the condensate at $T=0$
and compare it to Fig.~\ref{fig:sigmaall}.  There are modern ways based on Ref.~\cite{Giusti:1998wy},
but maybe a quicker (though dirtier) way is to use the  Gell Mann, Oakes, Renner relation, Eq.~\ref{eq:gmor}, using a single elimination jackknife from separate fits to the AWI quark mass,
the decay constant, and the pseudoscalar mass.
This will allow for a couple of checks: first, what are the lattice spacing effects in the data? and second, will it let us bracket the crossover region in temperature?

\begin{figure}
\begin{center}
\includegraphics[width=0.9\textwidth,clip]{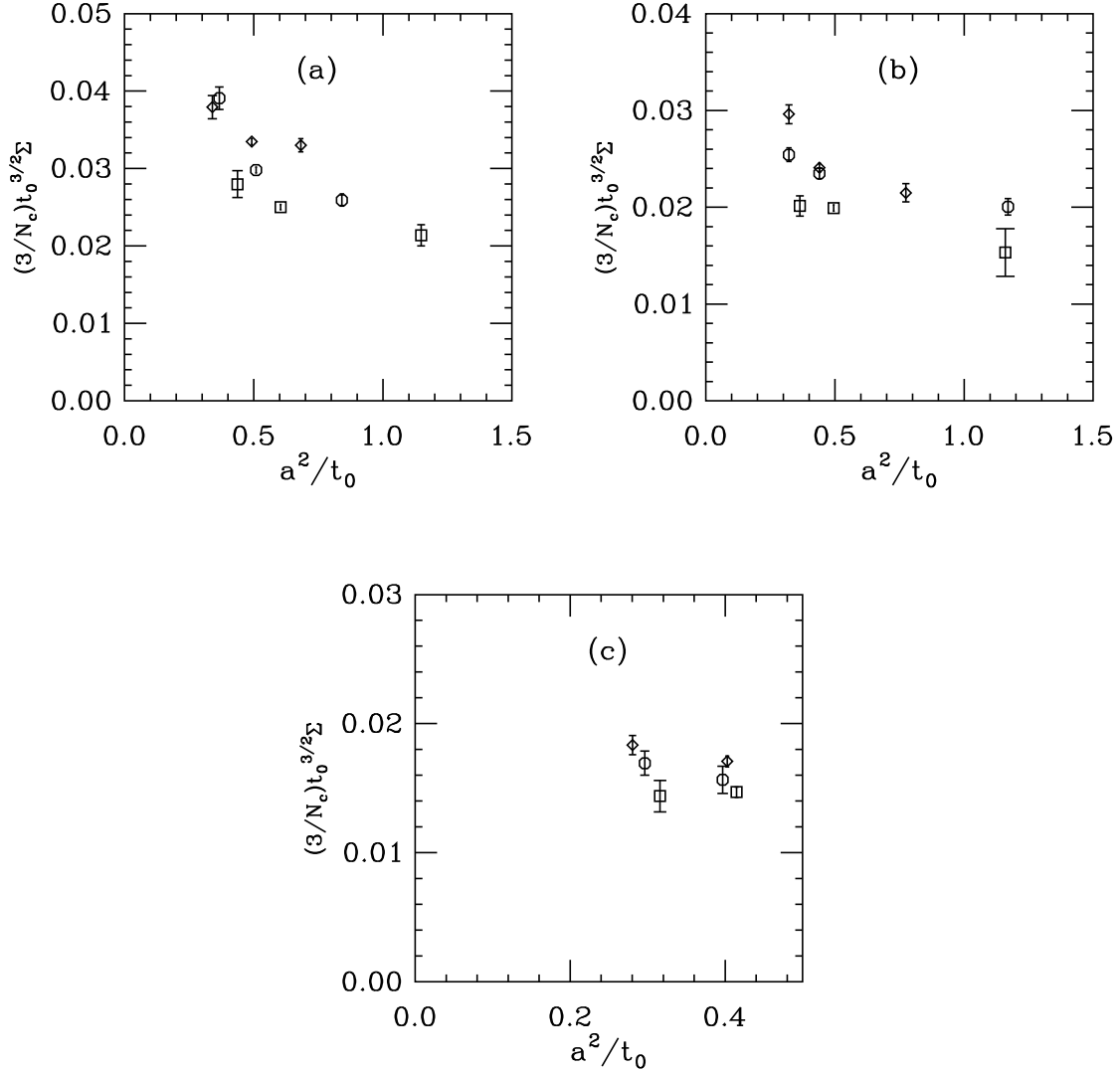}
\end{center}
\caption{Condensate from the Gell-Mann - Oakes - Renner relation versus  $a^2/t_0$.
Squares, octagons, and diamonds label $N_c=3$, 4, and 5.
(a)  $(m_{PS}/m_V)^2 \sim 0.63$; 
(b) $(m_{PS}/m_V)^2 \sim 0.5$;
(c) $(m_{PS}/m_V)^2 \sim 0.25$.
\label{fig:t032pbpvs1t0}}
\end{figure}

Fig.~\ref{fig:t032pbpvs1t0}  shows $\Sigma(m)$ plotted as a function of $a^2/t_0$ for the three collections of bare parameters used in
 the $(m_{PS}/m_V)^2=0.63$,
0.5, and 0.25 sets of finite temperature. The different plotting symbols correspond to different $N_c$ values.
Lattice artifacts are clearly present [though the reader is cautioned again, the data sets at an individual $N_c$
are not matched very well in quark mass and some of the variation may be due to intrinsic fermion mass dependence in $\Sigma(m)$].
With $\sqrt{t_0}$ a nominal 0.15 fm, the deviation of the data at strong coupling from its value at weak coupling is not surprising.

Comparing Figs.~\ref{fig:sigmaall} and \ref{fig:t032pbpvs1t0},
the zero temperature condensate appears (rather noisily) to be $\frac{3}{N_c} t_0^{3/2} \Sigma \sim0.03-0.04$ at $r=0.63$,
0.02 at $r=0.5$ and 0.015 at $r=0.25$. The results in Fig.~\ref{fig:sigmaall} seem to be plausible, in the sense that they fall to roughly the negative of the zero temperature condensate.

\begin{figure}
\begin{center}
\includegraphics[width=0.9\textwidth,clip]{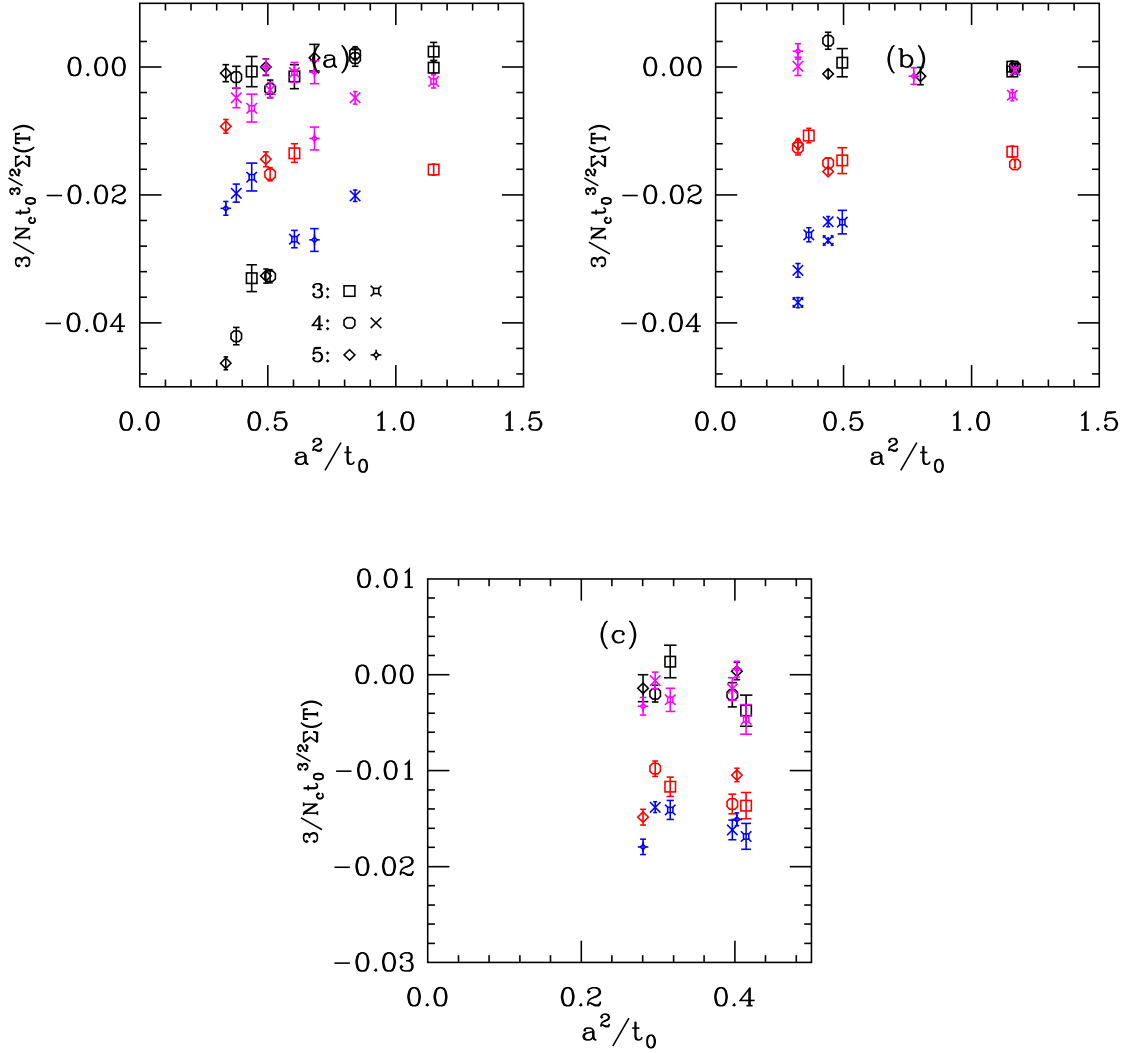}
\end{center}
\caption{Temperature dependent condensate versus  $a^2/t_0$.
Two sets of plotting symbols, shown in panel (a), label $N_c=3$, 4, and 5.
The panels are
(a)  $(m_{PS}/m_V)^2 \sim 0.63$; 
(b) $(m_{PS}/m_V)^2 \sim 0.5$;
(c) $(m_{PS}/m_V)^2 \sim 0.25$.
In panel (a) the data  proceeding down from the top are
(1) $\sqrt{t_0}T< 0.11$  in black with the first set of symbols,
(2) $\sqrt{t_0}T\sim 0.18$ in purple with the second set of symbols,
(3) $\sqrt{t_0}T\sim 0.21$ in red with the first set of symbols,
(4) $\sqrt{t_0}T\sim 0.25$ in blue with the second set of symbols,
(5) $\sqrt{t_0}T > 0.35$ in black with the first set of symbols.
In panel (b), proceeding down from the top are
(1) $\sqrt{t_0}T < 0.11$  in black with the first set of symbols,
(2)  $\sqrt{t_0}T\sim 0.15$  in purple with the second set of symbols,
(3) $\sqrt{t_0}T\sim 0.23$ in red with the first set of symbols,
(4) $\sqrt{t_0}T > 0.35$ in blue with the second set of symbols.
In panel (c) again from the top down are
(1) $\sqrt{t_0}T\sim 0.10$ in black with the first set of symbols,
(2) $\sqrt{t_0}T\sim 0.13$ in purple with the second set of symbols,
(3) $\sqrt{t_0}T\sim 0.20$ in red with the first set of symbols,
(4) $\sqrt{t_0}T > 0.25$ in blue with the second set of symbols.
\label{fig:t0sigvs1t0}}
\end{figure}

A further check before returning to physics should involve the finite temperature data sets themselves. I will break up the data for
the temperature-dependent condensate into bins in the temperature $\sqrt{t_0}T$ and plot it versus $a^2/t_0$.
This is shown in Fig.~\ref{fig:t0sigvs1t0}. The picture is a bit awkward to present; the data show  temperature dependence 
within  each bin in addition to scale dependence. I have color coded the various bins and used two sets of plotting symbols
for each $N_c$ value to separate the different $\sqrt{t_0}T$ values.

The highest temperature data sets are the ones with the largest lattice spacing dependence. These are uniformly
taken with $N_t=4$. Their collection was an attempt to get high above the crossover temperature to try to see
a flattening in $\Sigma(T)$ at high temperature, while still keeping to a large physical value of the simulation volume.
Their only use is for the largest temperature bins in  a calculation of $d\Sigma(T)/dT$ in Sec.~\ref{sec:bump}.
They are not used in the pictures of screening masses in Sec.~\ref{sec:screen}.

 \subsection{Looking for a peak \label{sec:bump}}

\begin{figure}
\begin{center}
\includegraphics[width=0.9\textwidth,clip]{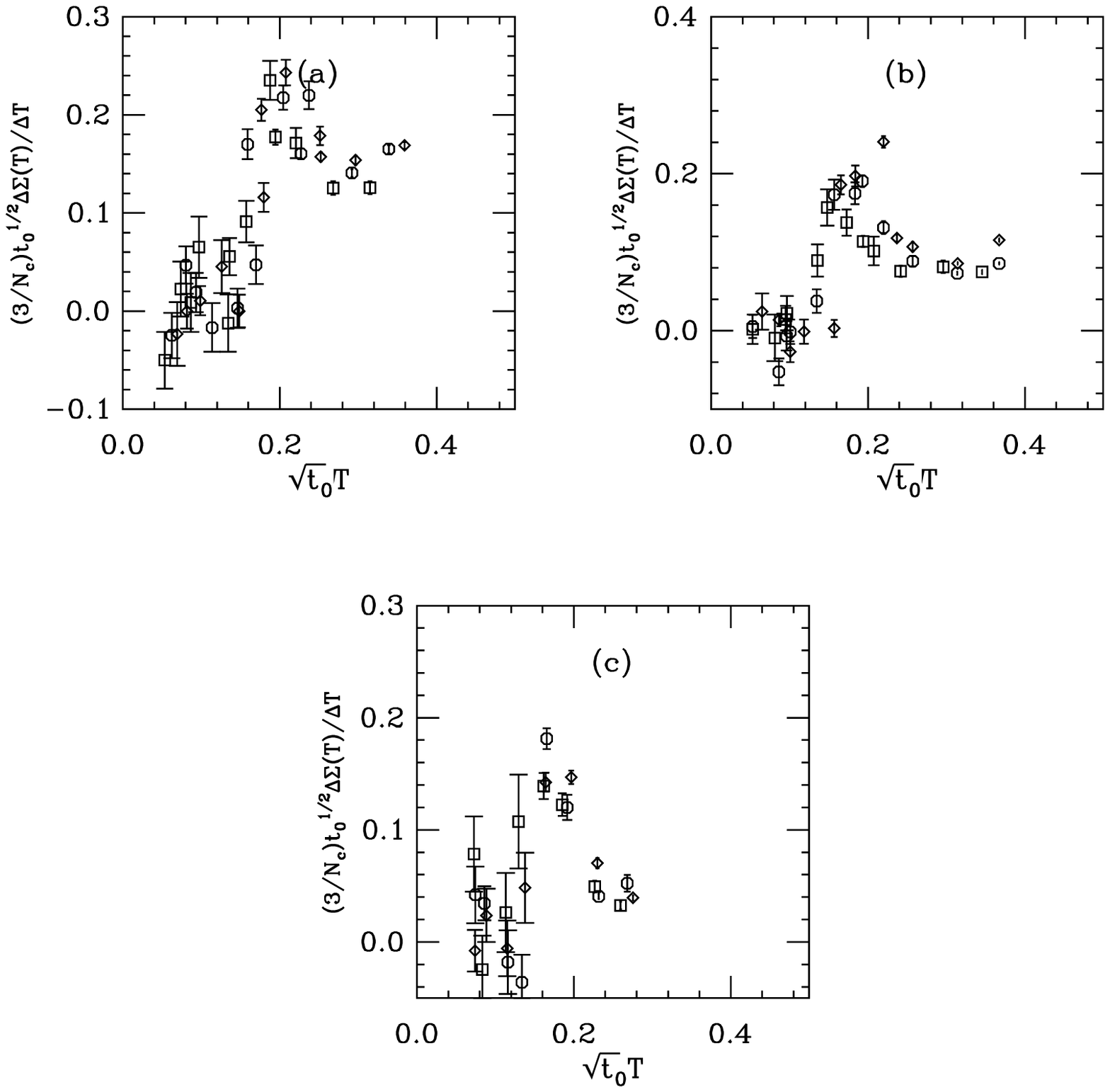}
\end{center}
\caption{$\Delta \Sigma(T)/\Delta T$, rescaled by $3/N_c$,
as a function of temperature, in appropriate units of $t_0$.
Squares, octagons, and diamonds label $N_c=3$, 4, and 5. 
(a)  $(m_{PS}/m_V)^2 \sim 0.63$; 
(b) $(m_{PS}/m_V)^2 \sim 0.5$;
(c) $(m_{PS}/m_V)^2 \sim 0.25$.
\label{fig:slopeall}}
\end{figure}

It would be better to have an observable with a peak, so I looked at two more quantities related to the condensate.
One of them produced a signal.
 It is the temperature derivative of $\Sigma(T)$, just taken from
the finite difference
\bee
\frac{\Delta \Sigma(T_m)}{\Delta T} = \frac{\Sigma(T_1) -\Sigma(T_2)}{T_1-T_2}
\label{eq:slope}
\ee
where  $T_m=(T_1+T_2)/2$.  The difference in Eq.~\ref{eq:slope} can just be taken from the
integrated pseudoscalar correlator at each value of $N_t$, without doing the $T=0$ subtraction. Of course,
only data sets at the same bare parameters can be used.
Fig.~\ref{fig:slopeall} shows this (rescaled by $N_c$ and the appropriate power of $t_0$).
The differences are (for example for $r=0.25$) $N_t=6-8$, 8-12, 12-16, and 16-32.

Fig.~\ref{fig:slopeall} shows a broad obvious feature at $\sqrt{t_0}T \sim 0.18$, pretty much independent of $(m_{PS}/m_V)^2$.
This corresponds to a temperature of about 225 MeV, an intermediate value between the quenched and physical transition points.
The figure certainly shows no difference between $N_c=3$, 4 and 5.

Attempts to refine this statement came to nothing. They were mostly based on doing fits to an arbitrary peaked function, a Gaussian,
\bee
y(x)= C_1\exp\left(-\frac{1}{2}\frac{(x-x_0)^2}{\sigma^2}\right)
\ee
with $x=\sqrt{t_0}T$. Fits to all data sets with the same $N_c$ value (for each choice of $(m_{PS}/m_V)^2$)
generally had poor chi-squared, probably due to lattice artifacts: the data span a wide range of $t_0$ values.
Fits to a single set of bare parameters fared better, though the issue is that there are only four values of $N_t$.
Some of the data sets (especially the ones at coarse lattice spacing) do not themselves include the peak
and then of course the fit fails immediately.

I also attempted to compute a susceptibility from the 
 time histories of $\Delta_{PP}(T)$. This was unsuccessful
so I do not report on it.

\section{Screening masses \label{sec:screen}}

Meson screening masses in the scalar, pseudoscalar, vector, and axial vector channels are taken
from two-point correlation functions extending in a spatial lattice direction. The temperature  dependent pseudoscalar
decay constant is also extracted from these spatial correlators.
Propagators are constructed  with composite boundary conditions to double the effective
length of the lattice \cite{Blum:2001xb,Aoki:2005ga,DeGrand:2007tx}.

In the low temperature, chirally broken phase, the spectroscopy
of screening masses should qualitatively resemble ordinary $T=0$ spectroscopy with a light pion and no degeneracies in the spectrum.
When chiral symmetry is restored parity partners (the pion and the scalar mesons, the vector and axial vector mesons)
should become degenerate, and all four states should become degenerate when $U(1)_A$ is restored.
 A naive expectation for a screening mass is that it
behaves  something like
\bee
m_H^2 = 4\left [ \left(\frac{\pi}{N_t}\right)^2 + m_q^2\right]
\label{eq:minmat}
\ee
where $\pi/N_t$ is the lowest nonzero Matsubara frequency associated with
antiperiodic boundary conditions in a lattice of temporal length $N_t$. Since $1/N_t=aT$,
this gives  $m_H = 2\pi T$ at high temperature.

\begin{figure}
\begin{center}
\includegraphics[width=0.9\textwidth,clip]{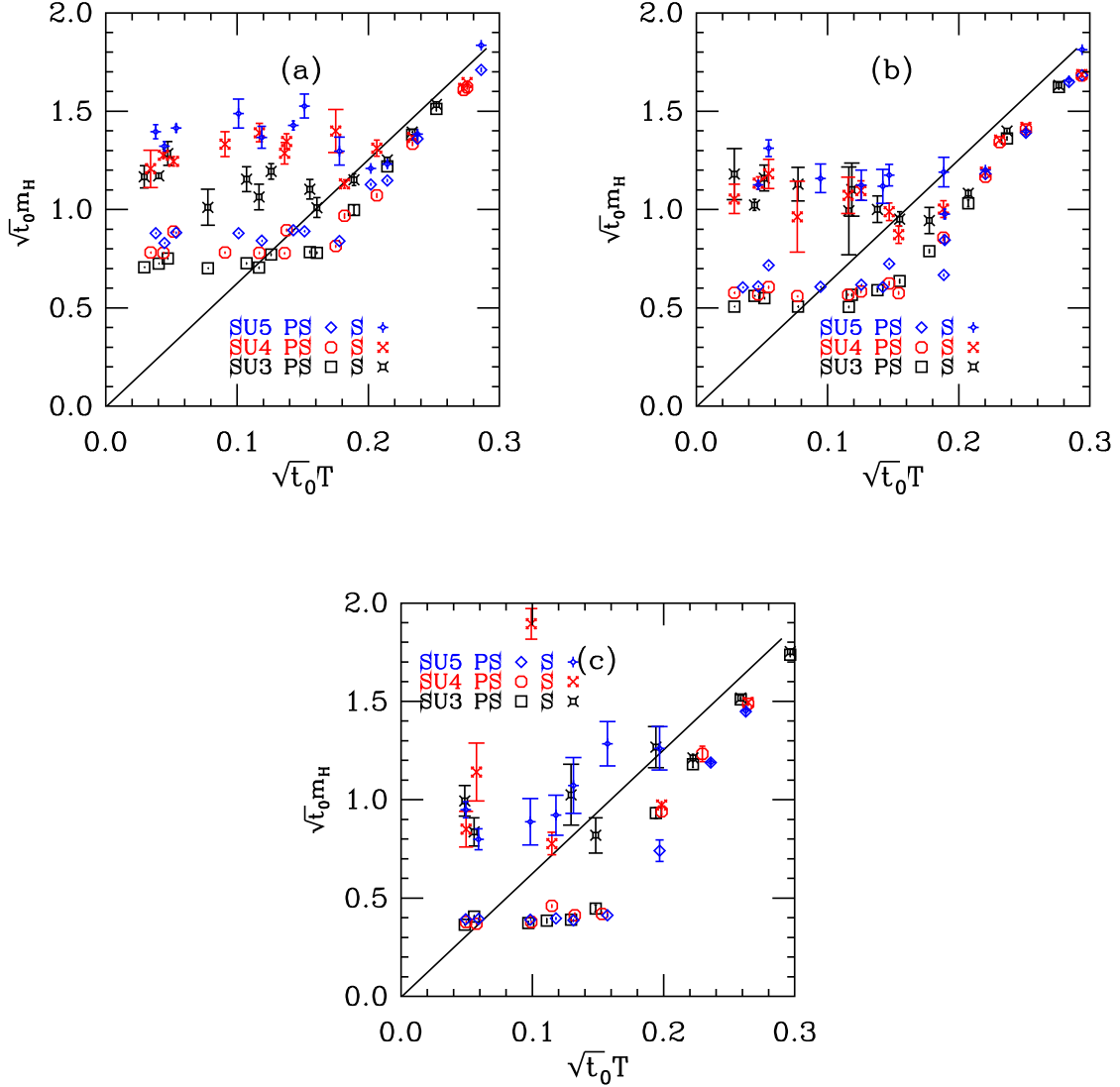}
\end{center}
\caption{Pseudoscalar and scalar meson  screening masses ($\sqrt{t_0}m_{PS}$ and $\sqrt{t_0}m_{S}$)
versus $\sqrt{t_0} T$. (a) $m_{PS}/m_V=0.63$, (b) $m_{PS}/m_V=0.5$ (c) $m_{PS}/m_V=0.25$.
  The panel shows the
plotting symbols with PS and S for pseudoscalar and scalar mesons, respectively. The 
$SU(3)$ results are shown in black, $SU(4)$ in red,  and $SU(5)$ in blue.
 The line is just $m_H=2\pi T$.
\label{fig:mpima0all}}
\end{figure}

\begin{figure}
\begin{center}
\includegraphics[width=0.9\textwidth,clip]{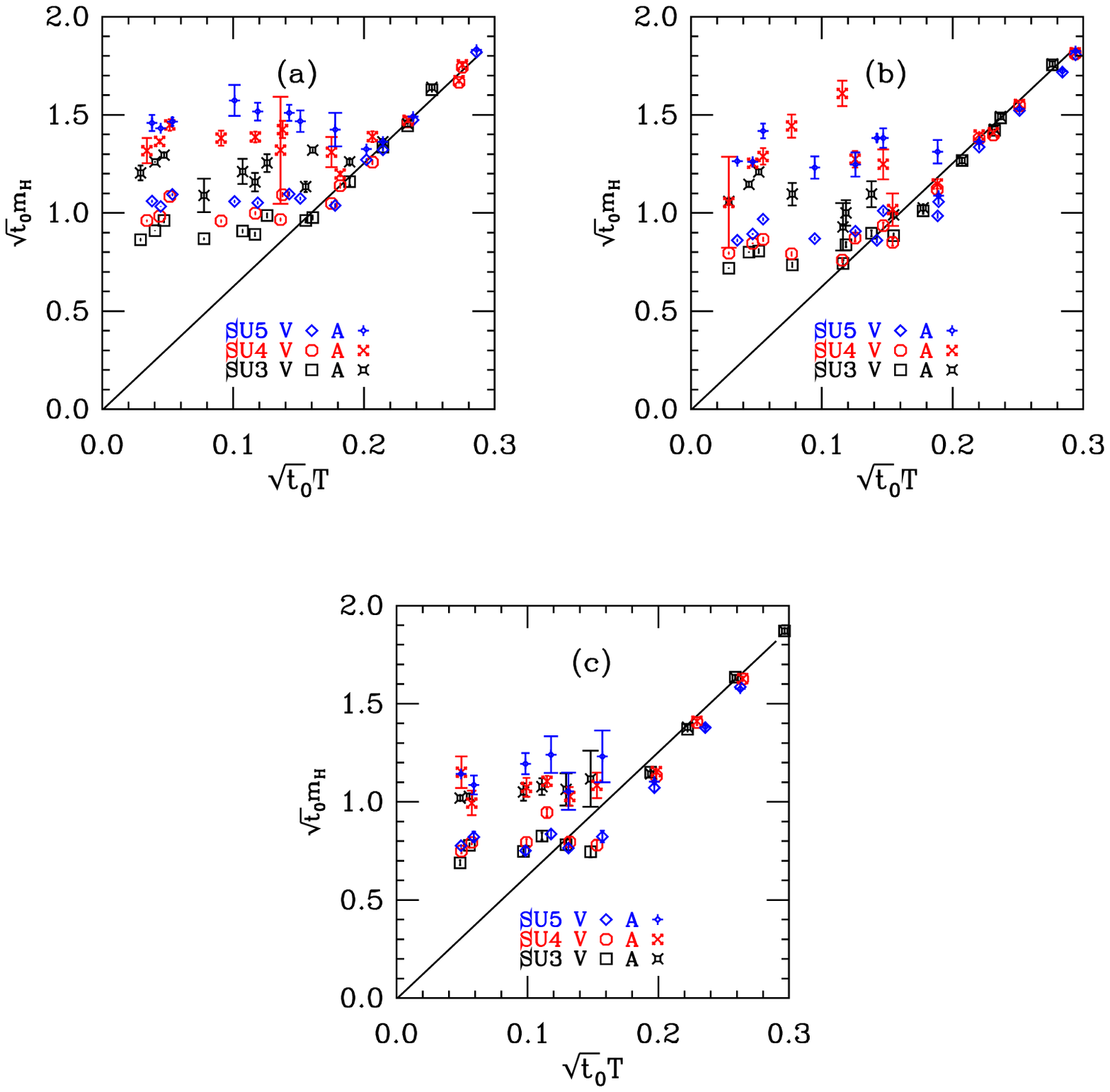}
\end{center}
\caption{Vector and axial vector meson screening mass ($\sqrt{t_0}m_{V}$ and $\sqrt{t_0}m_A$)
versus $\sqrt{t_0} T$. (a) $m_{PS}/m_V=0.63$, (b) $m_{PS}/m_V=0.5$ (c) $m_{PS}/m_V=0.25$.
 The panel shows the
plotting symbols with V and A for vector and axial vector meson masses respectively. The
$SU(3)$ results are shown in black, $SU(4)$ in red, and $SU(5)$ in blue.
 The line is just $m_H=2\pi T$.
\label{fig:mrhoma1all}}
\end{figure}

Results for screening masses  (scaled by $\sqrt{t_0}$) are displayed in 
Fig.~\ref{fig:mpima0all} 
for the pseudoscalar and scalar states
and Fig.~\ref{fig:mrhoma1all}
for the vector an axial vector states, 
They show the expected behavior.
(The diagonal lines are just $y= 2\pi T$, the small $m_q$ (or large $T$) limit of Eq.~\ref{eq:minmat}.)
The very dirty signals for the $a_0$ (scalar) and $a_1$ (axial vector) mesons in the chirally broken phase are also  expected.
There, the pseudoscalar meson is light and the noise to signal ratio  $\sigma(t)/C(t) \sim \exp((m_H-m_\pi))t$   
is exponentially bad  \cite{Lepage:1989hd,Parisi:1983ae}; in the chirally restored phase all states are massive
and the noise to signal ratio improves.

 The pseudoscalar and vector screening masses make a clear transition from a temperature independent
  value at low $T$ to linear behavior at high $T$. This suggests that linear fits $O=c_1+c_2 T$ 
  should show better quality (smaller chi-squared) when fits keeping
 only data in one phase are included, and $c_2$ would be much larger in the higher $T$ phase.
 The fit would deteriorate when points in both phases were included. This would determine a crossover temperature.
 
 The transition can be seen in individual data sets, and no fit is needed to see it, just a table of the mass values.
 In most sets the change in slope occurs in the middle of the data set, with (typically) two or three masses with nearly
 the same value at low temperature and the remaining higher temperature masses rising linearly with temperature.
 Data set by data set, one can identify a $\sqrt{t_0}T_{low}$ and a $\sqrt{t_0}T_{high}$ where masses remain unchanged for $T\le T_{low}$
 and rise linearly for $\sqrt{t_0}T\ge \sqrt{t_0}T_{high}$.  Generally,
 $\sqrt{t_0}T_{low}$ and  $\sqrt{t_0}T_{high}$ are consistent between the pseudoscalar and vector mass sets.
 The individual data sets give a relatively wide interval between $\sqrt{t_0}T_{low}$ and $\sqrt{t_0}T_{high}$ simply because
 $T=1/(aN_t)$ and the values of $N_t$ are small.
 
 Fits to all mass values at each $N_c$ and $(m_{PS}/m_V)^2$ value also show  jumps in chi-squared when a fit including
 high temperature points extends too low, or fits to low temperature points extend too high.
  The issue is that even with data in one phase, the chi-squared tended to be unacceptably large
 due to the $a$ dependent variation in the masses.   The best one can say is that the crossover region is in the range
 $\sqrt{t_0}T_c\sim 0.15-0.24$, which is not inconsistent with the location of the broad peak in $d\Sigma/dT$.

\begin{figure}
\begin{center}
\includegraphics[width=0.9\textwidth,clip]{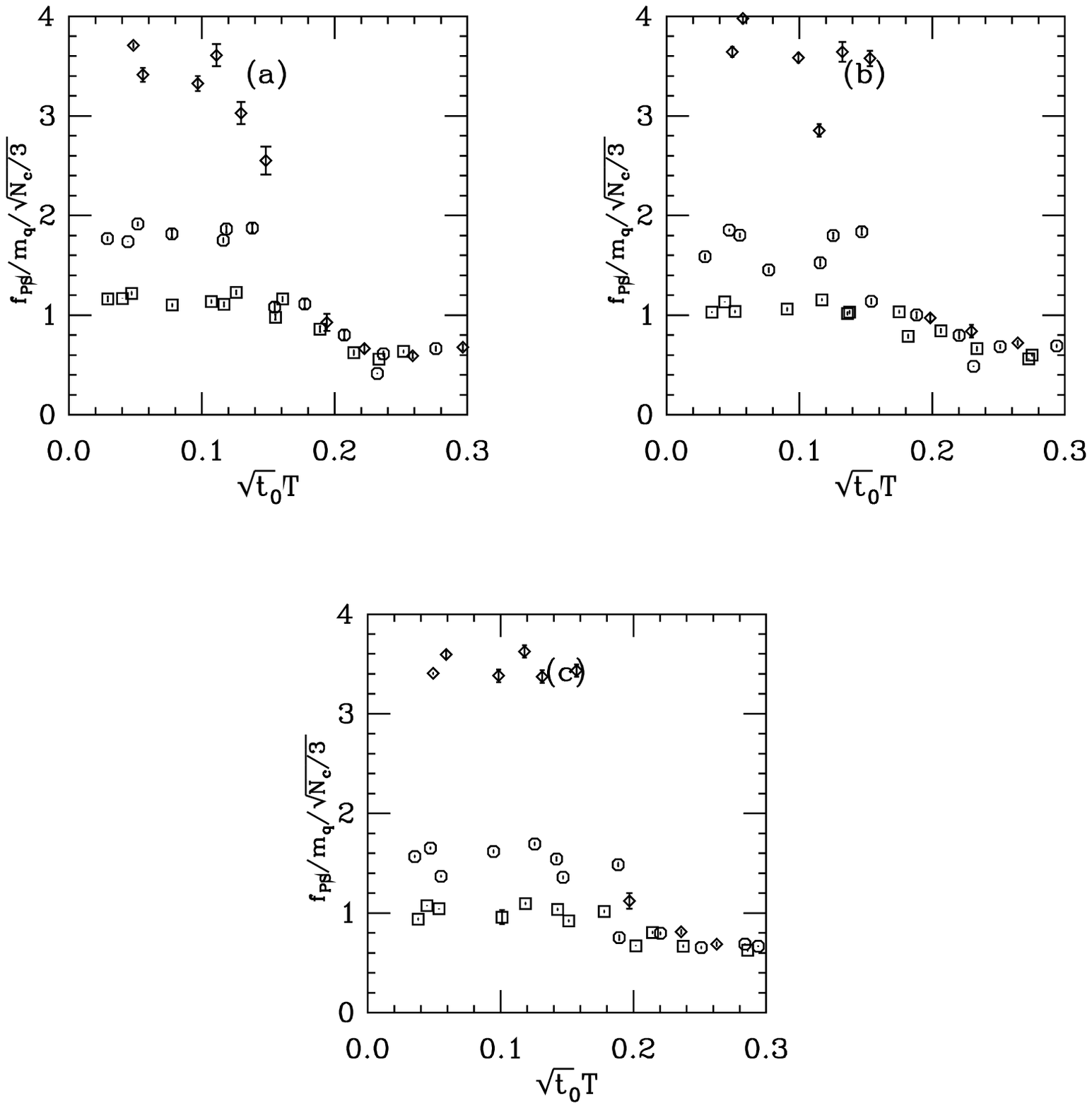}
\end{center}
\caption{Pseudoscalar decay constant divided by the AWI fermion mass  ($f_{PS}/m_q$) versus $\sqrt{t_0} T$. (a) $SU(3)$, (b) $SU(4)$, (c) $SU(5)$. The plotting symbols correspond to $(m_{PS}/m_V)^2=0.63$ for squares, 0.5 for octagons, and 0.25 for diamonds.
\label{fig:fpimqSUN}}
\end{figure}

The last screening quantity to present is the pseudoscalar decay constant. Here, the following alternative seems to
show a sharper result than plots of $f_{PS}$ versus temperature at fixed $r$.
In the chirally broken phase, $f_{PS}$ is nonvanishing at zero fermion mass (due to spontaneous breaking of chiral symmetry) and rises
modestly with fermion mass (due to  to explicit breaking of chiral symmetry from the fermion mass).
In the chirally restored phase there is only explicit symmetry breaking, and $f_{PS}$ should be proportional to $m_q$.
A plot of $f_{PS}/m_q$ will show collapse to a common value when that situation occurs.
Fig.~\ref{fig:fpimqSUN} shows that behavior. I have scaled $f_{PS}$ by $\sqrt{3/N_c}$.
Here the different plotting symbols in each panel represent different values of
$(m_{PS}/m_V)^2$ with the diamonds representing the lightest fermion mass.

\section{Polyakov line and its susceptibility \label{sec:poly}}

It is well known that the Polyakov line is not a sensitive observable at small fermion mass,
but I present one picture for completeness.
Following Ref.~\cite{Lucini:2012wq}, I define the volume averaged Polyakov line as
\bee
l_P=\frac{1}{N_s^3 N_c} \sum_x \Tr \prod_{t=0}^{N_t-1} U_4(x,t)
\ee
and then the susceptibility is
\bee
\chi_P = N_s^3( \svev{|l_P|^2} - \svev{|l_P|}^2 )  .
\label{eq:suscept}
\ee
I measure these quantities from time histories of the average Polyakov line
 taking a jackknife average over the simulation run.
I have only measured the original Polyakov line, not  any smoothed one.

\begin{figure}
\begin{center}
\includegraphics[width=0.9\textwidth,clip]{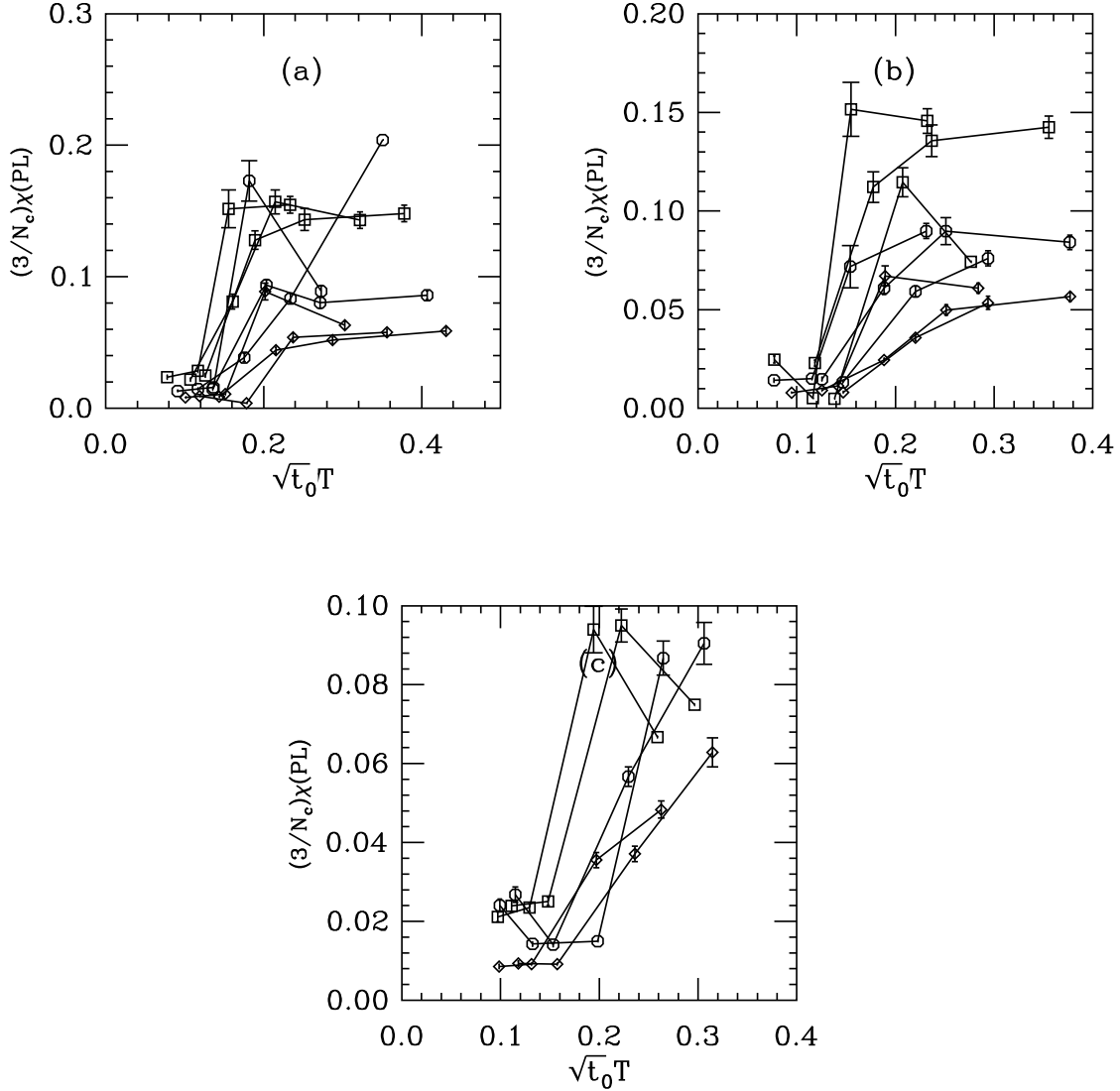}
\end{center}
\caption{Polyakov line susceptibility
as a function of temperature, scaled by $\sqrt{t_0}$.
Squares, octagons, and diamonds label $N_c=3$, 4, and 5. Lines connect simulation results from the same bare parameters.
(a)  $(m_{PS}/m_V)^2 \sim 0.63$; 
(b) $(m_{PS}/m_V)^2 \sim 0.5$;
(c) $(m_{PS}/m_V)^2 \sim 0.25$.
\label{fig:plscall}}
\end{figure}

The Polyakov line susceptibility is shown in Fig.~\ref{fig:plscall}. The bare Polyakov line is not a scaling quantity,
so the results for different $N_c$ values should not coincide. The only comment to make about these noisy figures is that the magnitude of
the susceptibility falls with the fermion mass, as seen  for $N_c=3$ \cite{Clarke:2019tzf}.

At small fermion masses, where one is closer
to the $m_q=0$ critical point, the $N_c=3$  Polyakov line and its derivatives do show structure
associated with the critical point \cite{Clarke:2020htu}.
Since my data were collected 
at heavier quark masses where everything is crossover, the Polyakov line was not a particularly useful observable.

Note that the shoulder in the Polyakov line susceptibility appears at $\sqrt{t_0}T\sim 0.15$ or so, about 
where the screening masses begin to take their high temperature functional form.

\section{Conclusions\label{sec:conclusions}}

The question this study was designed to address was whether the
finite temperature crossover behavior of $SU(N_c)$ gauge theories with $N_f=2$ flavors of fermions showed
different behavior as the number of colors was varied.
These simulations, with low statistics and carried out on small volumes, studied 
the temperature dependence of the condensate (as determined from the volume integral of the
pseudoscalar correlator) and of screening masses and the pseudoscalar decay constant.
A smooth crossover from 
a low temperature confining and chirally broken phase to a high temperature chirally restored and deconfined phase is observed.
This crossover behavior is of course not surprising given our extensive knowledge of $N_c=3$:
 the one new result is that the temperature dependence of these quantities (when compared at fixed values of $m_{PS}/m_V$)
shows no observable dependence on $N_c$.

It was not possible to determine a crossover temperature (to the extent that such a quantity makes sense when
the crossover is smooth) but,  with some plausible assumptions, it appears to be someplace between the known 
$SU(3)$ result with physical quark masses ($\sqrt{t_0}T\sim 0.11$) and the large $N_c$ pure gauge result ($\sqrt{t_0}T_c\sim 0.26$).

Of course, one can also make comparisons with earlier work. Refs.~\cite{AliKhan:2001ek,Ejiri:2009hq} present $SU(3)$ results for
$T_c/m_V$ as a function of $m_{PS}/m_V$. With a nominal crossover temperature $\sqrt{t_0}T_c\sim 0.18$ and $m_V$ taken from Table~\ref{tab:datasets},
$T_c/m_V \sim 0.20$, 0.23, 0.25 at $r=0.63$, 0.5, 0.25, in reasonably good agreement with   Refs.~\cite{AliKhan:2001ek,Ejiri:2009hq}.
Ref.~\cite{Bornyakov:2009qh}  shows a plot of $r_0 T_c$ versus $r_0 m_{PS}$ where $r_0$ is the Sommer parameter \cite{Sommer:1993ce},
$r_0/\sqrt{t_0}\sim 3$.  This conversion gives $r_0 T_c \sim 0.54$ at $r_0 m_{PS}=2.2$, 1.6 and 1.1 at $r=0.63$, 0.5, 0.25,
to be compared with $r_0 T_c =0.48-0.6$ from Ref.~\cite{Bornyakov:2009qh} for the same range of $r_0 m_{PS}$ values.

Of the three scenarios described in the Introduction, the first two seem to be disfavored. No evidence was seen for
any first order behavior at large mass at any $N_c$ studied. Of course, it could be that $N_c=5$ is still not ``large $N_c$''
from the point of view of thermodynamics. Scenario two  suggests crossover behavior, which is seen,
but  the scenario that $T_c\propto f_{PS}$
is disfavored  (of course, over the range of fermion masses studied)  because 
the crossover temperature shows no obvious $N_c$ separation despite the (known) variation of the
pseudoscalar decay constant with respect to $N_c$. 
As far as the third scenario, lattice simulations show that the lowest
part of the meson spectrum show little $N_c$ dependence. The third scenario assumes that the two results I have presented
-- a common spectrum and a common crossover behavior -- are correlated.
Of course, the scenario asks that the correlation persists high in the meson spectrum (that  the density of states is as given by
Hagedorn) and lattice simulations say nothing about that.

The results in this paper are rather poor quality, but at the mass values studied, everything is smooth.
The situation at very low mass and very high mass remains open.
I think that to do more work in this area, one should not continue to use Wilson-clover fermions -- at least, not in pilot tests.
The importance of chiral symmetry, the
 need to subtract results from finite and zero temperature simulations, and the fierce scaling of the cost with
simulation volume of thermodynamical observables argue in favor of using
some modern version of staggered fermions for dedicated simulations.

The obvious next target in future studies of large $N_c$ systems at finite temperature could be a more direct
attack on thermodynamic observables -- the internal energy, pressure, speed of sound, and related quantities.
The pure gauge calculations of Panero \cite{Panero:2009tv} are an  inspiration, and  they have been extensively cited
in the phenomenological and gravitational duality literature of QCD thermodynamics. But we know
that the critical behavior of $N_c=3$ with dynamical fermions is different from the behavior of
pure gauge systems. Does this matter, for the questions which interest researchers in these areas?
If so, numerical simulation which includes dynamical fermions might be a worthwhile project.

\begin{acknowledgments}
Daniel Hackett participated in the earliest stages of this project.
I am grateful to Rob Pisarski for a conversation about large $N_c$ expectations for QCD thermodynamics.
My computer code is based on the publicly available package of the
 MILC collaboration~\cite{MILC}. The version I use was originally developed by Y.~Shamir and
 B.~Svetitsky.
This material is based upon work supported by the U.S. Department of Energy, Office of Science, Office of
High Energy Physics under Award Number DE-SC-0010005.
Some of the computations for this work were also carried out with resources provided by the USQCD
Collaboration, which is funded
by the Office of Science of the U.S.\ Department of Energy
using the resources of the Fermi National Accelerator Laboratory (Fermilab), a U.S.
Department of Energy, Office of Science, HEP User Facility. Fermilab is managed by
 Fermi Research Alliance, LLC (FRA), acting under Contract No. DE- AC02-07CH11359.
\end{acknowledgments}

\begin{table}
\begin{tabular}{c c c c c c c c c c}
\hline
$N_c$ & $\beta$ & $\kappa$ & $\kappa_c$ & $am_q$ & $a\,m_{PS}$  & $a\,m_V$ & $(m_{PS}/m_V)^2$ & $t_0/a^2$ &   N \\
\hline
$r \sim 0.63$  &  & & & & & &    & &\\
\hline
3 &5.3 & 0.1250 & 0.12923 & 0.153 & 0.757(1)& 0.925(5)& 0.670(7)& 0.871(3)  & 30 \\
3 &5.4 & 0.1250 & 0.12838 & 0.107 & 0.563(1)& 0.707(1)& 0.634(3)& 1.657(3)  & 400 \\
3 &5.45 & 0.1250 & 0.12795 & 0.093 & 0.497(2)& 0.636(5)& 0.611(11)& 2.284(21)  & 30 \\
4 & 10.1 & 0.1240 & 0.12794 & 0.150 & 0.716(2)& 0.881(3)& 0.661(6)& 1.189(7) & 30 \\ 
4 & 10.2 & 0.1245 & 0.12792 & 0.108 & 0.556(1)& 0.701(1)& 0.629(3)& 1.966(3)  & 190 \\
4 & 10.3 & 0.1240  & 0.12759 & 0.112 & 0.537(2)& 0.657(3)& 0.668(8)& 2.727(18)  & 50 \\
5 & 16.3 & 0.1230 & 0.12795  & 0.162 & 0.726(1)& 0.875(3)& 0.688(5)& 1.467(7)  & 30 \\
5 & 16.4 & 0.1240 & 0.12785 & 0.119 & 0.582(1)& 0.725(1)& 0.644(3)& 2.030(2)  & 190 \\
5 & 16.6 & 0.1240  & 0.12740 & 0.106 & 0.514(2)& 0.638(2)& 0.649(6)& 2.942(14)  & 50 \\
\hline
$r \sim 0.5$  &  & & & & & &    & &\\
\hline
3 & 5.25 & 0.1280 & 0.12964 & 0.080 & 0.545(2)& 0.773(4)& 0.497(6)& 0.863(2) & 30 \\ 
3 & 5.4 & 0.1265 & 0.12838 & 0.058 & 0.395(1)& 0.563(2)& 0.492(4)& 2.019(5)  & 400 \\
3 & 5.45 & 0.1265  & 0.12795  & 0.044 & 0.331(2)& 0.486(6)& 0.464(13)& 2.747(17)  & 91 \\
4 & 10.0 & 0.1270 & 0.12926  & 0.104 & 0.623(3)& 0.860(6)& 0.525(9)& 0.855(2) & 40 \\ 
4 & 10.2 & 0.1262 & 0.12792  & 0.054 & 0.377(1)& 0.561(2)& 0.452(4)& 2.270(4)  & 190 \\
4 & 10.3 & 0.1260 & 0.12759 & 0.049 & 0.343(3)& 0.491(5)& 0.488(13)& 3.106(17)  & 90 \\
5 & 16.2 & 0.1260 & 0.12853 & 0.087 & 0.531(1)& 0.757(4)& 0.492(6)& 1.290(7)  & 30 \\
5 & 16.4 & 0.1258  & 0.12785 & 0.063 & 0.404(1)& 0.592(2)& 0.466(4)& 2.272(5)  & 190 \\
5 & 16.6 & 0.1252  & 0.12740 & 0.069 & 0.406(2)& 0.549(3)& 0.547(8)& 3.108(19)  & 40 \\
\hline
$r \sim 0.25$  &  & & & & & &    & &\\
\hline
3 &5.4 & 0.1276 & 0.12838  & 0.021 & 0.234(2)& 0.444(8)& 0.278(11)& 2.412(7) & 400 \\
3 &5.45 & 0.1273 & 0.12795 & 0.018 & 0.228(4)& 0.437(6)& 0.272(12)& 3.164(23) & 90 \\
4 & 10.2 & 0.1272 & 0.12792  & 0.022 & 0.238(2)& 0.472(8)& 0.254(10)& 2.520(20) & 101 \\ 
4 & 10.3 & 0.1270  & 0.12759 & 0.017 & 0.201(3)& 0.432(6)& 0.216(9)& 3.375(21)  & 90 \\
5 & 16.4 & 0.1270 & 0.12785  & 0.025 & 0.248(1)& 0.493(3)& 0.253(4)& 2.483(6) & 210 \\ 
5 & 16.6 & 0.1268 & 0.12740 & 0.019 & 0.208(2)& 0.435(14)& 0.229(16)& 3.565(23)  & 50 \\
\hline
 \end{tabular}
\caption{ Zero temperature data sets from $16^3 \times 32$ volumes.
(The $SU(5)$ $\beta=16.6$ $\kappa=0.1268$ set is $16^3\times 48$.). The last column gives the number of 
measurement lattices in the set. $\kappa_c$ is the value of the hopping parameter where the Axial Ward Identity 
fermion mass vanishes.
It is needed to perform tadpole renormalization of condensate - related quantities.
\label{tab:datasets}}
\end{table}

\begin{table}
\begin{tabular}{c c c c c c c c}
\hline
$N_c$ & $\beta$ & $\kappa$   & $N_t=4$ &  $N_t=6$ & $N_t=8$ & $N_t=12$ & $N_t=32$ \\
\hline
3  & 5.3 & 0.1250 &  16.64(2) & 18.11(6)& 18.34(5) & 18.61(12) & 18.35(9) \\ 
3  & 5.4 & 0.1250 &  16.49(2) & 17.30(4)& 18.06(5) & 18.02(8) & 18.11(8) \\ 
3  & 5.45 & 0.1250 &  16.45(2) & 17.14(3)& 17.61(3) & 17.86(5) & 17.89(9) \\ 
4  & 10.1 & 0.1240 &  22.03(2) & 23.44(4)& 24.08(5) & 24.01(9) & 23.88(8) \\ 
4  & 10.2 & 0.1245 &  21.93(2) & 22.91(2)& 23.72(4) & 23.73(6) & 23.94(6) \\ 
4  & 10.3 & 0.1240 &  21.84(1) & 22.68(2)& 23.24(3) & 23.36(4) & 23.42(5) \\ 
5  & 16.3 & 0.1230 &  27.40(2) & 28.58(3)& 29.35(3) & 29.52(8) & 29.41(13) \\ 
5  & 16.4 & 0.1240 &  27.41(2) & 28.60(3)& 29.54(4) & 29.54(5) & 29.54(7) \\ 
5  & 16.6 & 0.1240 &  27.26(1) & 28.28(2)& 28.82(2) & 29.17(4) & 29.21(4) \\ 
 \hline
\end{tabular}
\caption{ The (bare, lattice regulated) quantity $\hat \Delta_{PP}(N_t)$ as 
defined in Eq.~\protect{\ref{eq:delta}} for $r \sim 0.63$. Lattice volumes are $16^3 \times N_t$.
\label{tab:sig63}}
\end{table}

\begin{table}
\begin{tabular}{c c c c c c c c }
\hline
$N_c$ & $\beta$ & $\kappa$ & $N_t=4$ &  $N_t=6$ & $N_t=8$ & $N_t=12$ & $N_t=32$ \\
\hline
3  & 5.25 & 0.1280 & 16.90(9)     & 18.94(9) & 20.12(16)& 19.94(12) & 19.96(18)  \\ 
3  & 5.4 & 0.1265 &  16.58(2)   & 17.43(8) & 17.96(5)& 18.78(11) & 18.72(16)  \\ 
3  & 5.45 & 0.1265 & 16.49(2)    & 17.26(2) & 17.65(3)& 18.36(8) & 18.45(8)  \\ 
4  & 10.0 & 0.1270 & 22.44(9)    & 25.88(10) & 26.22(9)& 26.16(14) & 26.00(9)  \\ 
4  & 10.2 & 0.1262 & 22.10(2)    & 23.08(2) & 23.96(5)& 25.12(12) & 24.68(8)  \\ 
4  & 10.3 & 0.1260 & 21.99(1)    & 22.92(2) & 23.41(3)& 24.35(7) & 24.34(8)  \\ 
5  & 16.2 & 0.1260 & 27.79(2)    & 29.87(7) & 31.51(7)& 31.51(10) & 31.77(23)  \\ 
5  & 16.4 & 0.1258 & 27.61(2)    & 28.82(2) & 30.52(5)& 30.54(6) & 30.66(4)  \\ 
5  & 16.6 & 0.1252 & 27.40(1)    & 28.50(2) & 29.01(2)& 29.95(6) & 29.79(5)  \\ 
 \hline
 \end{tabular}
\caption{ The (bare, lattice regulated) quantity $\hat \Delta_{PP}(N_t)$ at $r\sim 0.5$.
 Lattice volumes are $16^3 \times N_t$.
\label{tab:sig5}}
\end{table}

\begin{table}
\begin{tabular}{c c c c c c c c}
\hline
$N_c$ & $\beta$ & $\kappa$   & $N_t=6$ &  $N_t=8$ & $N_t=12$ & $N_t=16$ & $N_t=32$ \\
\hline
3  & 5.4 & 0.1276 &  17.43(3) & 18.07(6)& 19.87(15) & 20.05(18) & 20.80(27) \\ 
3  & 5.45 & 0.1273 &  17.32(2) & 17.69(3)& 19.08(11) & 19.69(21) & 19.48(15) \\ 
4  & 10.2 & 0.1272 &  23.25(2) & 23.88(3)& 26.69(14) & 26.55(17) & 27.04(24) \\ 
4  & 10.3 & 0.1270 &  23.00(2) & 23.79(11)& 25.60(13) & 25.33(13) & 25.72(11) \\ 
5  & 16.4 & 0.1270 &  29.00(3) & 30.21(7)& 33.10(13) & 33.05(17) & 32.95(17) \\ 
5  & 16.6 & 0.1268 &  28.66(2) & 29.28(3)& 31.59(9) & 31.97(23) & 32.25(16) \\ 
 \hline
\end{tabular}
\caption{ The (bare, lattice regulated) quantity $\hat \Delta_{PP}(N_t)$ at
 $r\sim 0.25$. Lattice volumes are $16^3 \times N_t$.
\label{tab:sig25}}
\end{table}



\begin{thebibliography}{99}



\bibitem{tHooft:1973alw}
  G.~'t Hooft,
  Nucl.\ Phys.\ B {\bf 72}, 461 (1974).
  doi:10.1016/0550-3213(74)90154-0

\bibitem{tHooft:1974pnl}
  G.~'t Hooft,
  Nucl.\ Phys.\ B {\bf 75}, 461 (1974).
  doi:10.1016/0550-3213(74)90088-1
  
  
\bibitem{Witten:1979kh}
E.~Witten,
Nucl. Phys. B \textbf{160}, 57-115 (1979)
doi:10.1016/0550-3213(79)90232-3



\bibitem{Lucini:2012gg}
  B.~Lucini and M.~Panero,
  Phys.\ Rept.\  {\bf 526}, 93 (2013)
  doi:10.1016/j.physrep.2013.01.001
  [arXiv:1210.4997 [hep-th]].
  
  
  
\bibitem{GarciaPerez:2020gnf}
  M.~Garcia Perez,
Proc. Sci., LATTICE2019 (2020) 276
  arXiv:2001.10859 [hep-lat].
  
\bibitem{Hernandez:2020tbc}
P.~Hern\'andez and F.~Romero-L\'opez,
Eur. Phys. J. A \textbf{57}, no.2, 52 (2021)
doi:10.1140/epja/s10050-021-00374-2
[arXiv:2012.03331 [hep-lat]].


\bibitem{Donini:2016lwz}
A.~Donini, P.~Hernández, C.~Pena and F.~Romero-López,
Phys. Rev. D \textbf{94}, no.11, 114511 (2016)
doi:10.1103/PhysRevD.94.114511
[arXiv:1607.03262 [hep-ph]].


\bibitem{Donini:2020qfu}
A.~Donini, P.~Hern\'andez, C.~Pena and F.~Romero-L\'opez,
Eur. Phys. J. C \textbf{80}, no.7, 638 (2020)
doi:10.1140/epjc/s10052-020-8192-3
[arXiv:2003.10293 [hep-lat]].

\bibitem{Hernandez:2019qed}
P.~Hern\'andez, C.~Pena and F.~Romero-L\'opez,
Eur. Phys. J. C \textbf{79}, no.10, 865 (2019)
doi:10.1140/epjc/s10052-019-7395-y
[arXiv:1907.11511 [hep-lat]].

\bibitem{DeGrand:2020utq}
T.~DeGrand,
Phys. Rev. D \textbf{101}, no.11, 114509 (2020)
doi:10.1103/PhysRevD.101.114509
[arXiv:2004.09649 [hep-lat]].


\bibitem{Kaiser:2000gs} 
  R.~Kaiser and H.~Leutwyler,
  Eur.\ Phys.\ J.\ C {\bf 17}, 623 (2000)
  doi:10.1007/s100520000499
  [hep-ph/0007101].


\bibitem{Pisarski:1983ms} 
  R.~D.~Pisarski and F.~Wilczek,
  Phys.\ Rev.\ D {\bf 29}, 338 (1984).
  doi:10.1103/PhysRevD.29.338

\bibitem{Pisarski} Thanks to Rob Pisarski for discussions about this point.



\bibitem{Hagedorn:1968zz} 
  R.~Hagedorn,
  Nuovo Cim.\ A {\bf 56}, 1027 (1968).
  doi:10.1007/BF02751614



\bibitem{Cabibbo:1975ig} 
  N.~Cabibbo and G.~Parisi,
  Phys.\ Lett.\  {\bf 59B}, 67 (1975).
  doi:10.1016/0370-2693(75)90158-6


\bibitem{Lucini:2005vg} 
  B.~Lucini, M.~Teper and U.~Wenger,
  JHEP {\bf 0502}, 033 (2005)
  doi:10.1088/1126-6708/2005/02/033
  [hep-lat/0502003].

\bibitem{Lucini:2012wq} 
  B.~Lucini, A.~Rago and E.~Rinaldi,
  Phys.\ Lett.\ B {\bf 712}, 279 (2012)
  doi:10.1016/j.physletb.2012.04.070
  [arXiv:1202.6684 [hep-lat]].


\bibitem{Cuteri:2020yke}
F.~Cuteri, O.~Philipsen, A.~Sch\"on and A.~Sciarra,
Phys. Rev. D \textbf{103}, no.1, 014513 (2021)
doi:10.1103/PhysRevD.103.014513
[arXiv:2009.14033 [hep-lat]].

\bibitem{Hasenfratz:2001hp}
  A.~Hasenfratz and F.~Knechtli,
  Phys.\ Rev.\ D {\bf 64}, 034504 (2001).
  doi:10.1103/PhysRevD.64.034504
  [hep-lat/0103029].


\bibitem{Hasenfratz:2007rf}
  A.~Hasenfratz, R.~Hoffmann and S.~Schaefer,
  JHEP {\bf 0705}, 029 (2007).
  doi:10.1088/1126-6708/2007/05/029
  [hep-lat/0702028].




\bibitem{DeGrand:2012qa}
  T.~DeGrand, Y.~Shamir and B.~Svetitsky,
  Phys.\ Rev.\ D {\bf 85}, 074506 (2012).
  doi:10.1103/PhysRevD.85.074506
  [arXiv:1202.2675 [hep-lat]].

\bibitem{DeGrand:2016pur}
  T.~DeGrand and Y.~Liu,
  Phys.\ Rev.\ D {\bf 94}, no. 3, 034506 (2016)
  Erratum: [Phys.\ Rev.\ D {\bf 95}, no. 1, 019902 (2017)]
  doi:10.1103/PhysRevD.95.019902, 10.1103/PhysRevD.94.034506
  [arXiv:1606.01277 [hep-lat]].


\bibitem{Duane:1986iw}
  S.~Duane and J.~B.~Kogut,
  Nucl.\ Phys.\ B {\bf 275}, 398 (1986).
  doi:10.1016/0550-3213(86)90606-1

\bibitem{Duane:1985hz}
  S.~Duane and J.~B.~Kogut,
  Phys.\ Rev.\ Lett.\  {\bf 55}, 2774 (1985).
  doi:10.1103/PhysRevLett.55.2774

\bibitem{Gottlieb:1987mq}
  S.~A.~Gottlieb, W.~Liu, D.~Toussaint, R.~L.~Renken and R.~L.~Sugar,
  Phys.\ Rev.\ D {\bf 35}, 2531 (1987).
  doi:10.1103/PhysRevD.35.2531

\bibitem{Takaishi:2005tz}
  T.~Takaishi and P.~de Forcrand,
  Phys.\ Rev.\ E {\bf 73}, 036706 (2006).
  doi:10.1103/PhysRevE.73.036706
  [hep-lat/0505020].

\bibitem{Urbach:2005ji}
  C.~Urbach, K.~Jansen, A.~Shindler and U.~Wenger,
  Comput.\ Phys.\ Commun.\  {\bf 174}, 87 (2006).
  doi:10.1016/j.cpc.2005.08.006
  [hep-lat/0506011].


\bibitem{Hasenbusch:2001ne}
  M.~Hasenbusch,
  Phys.\ Lett.\ B {\bf 519}, 177 (2001).
  doi:10.1016/S0370-2693(01)01102-9
  [hep-lat/0107019].
  
\bibitem{other}
  R.~Narayanan and H.~Neuberger,
  JHEP {\bf 0603}, 064 (2006)
  doi:10.1088/1126-6708/2006/03/064
  [hep-th/0601210].
%
  M.~Luscher,
  Commun.\ Math.\ Phys.\  {\bf 293}, 899 (2010)
  doi:10.1007/s00220-009-0953-7
  [arXiv:0907.5491 [hep-lat]].

\bibitem{Luscher:2010iy}
  M.~L\"uscher,
  JHEP {\bf 1008}, 071 (2010)
  Erratum: [JHEP {\bf 1403}, 092 (2014)]
  doi:10.1007/JHEP08(2010)071, 10.1007/JHEP03(2014)092
  [arXiv:1006.4518 [hep-lat]].
  
  
\bibitem{Sommer:1993ce}
  R.~Sommer,
  Nucl.\ Phys.\ B {\bf 411}, 839 (1994).
  doi:10.1016/0550-3213(94)90473-1
  [hep-lat/9310022].


\bibitem{DeGrand:2012hd}
  T.~DeGrand,
  Phys.\ Rev.\ D {\bf 86}, 034508 (2012)
  doi:10.1103/PhysRevD.86.034508
  [arXiv:1205.0235 [hep-lat]].

  
  
\bibitem{Lottini:2013rfa}
  S.~Lottini [ALPHA Collaboration],
  PoS LATTICE {\bf 2013}, 315 (2014)
  doi:10.22323/1.187.0315
  [arXiv:1311.3081 [hep-lat]].



\bibitem{Bruno:2013gha}
  M.~Bruno {\it et al.} [ALPHA Collaboration],
  PoS LATTICE {\bf 2013}, 321 (2014)
  doi:10.22323/1.187.0321
  [arXiv:1311.5585 [hep-lat]].

\bibitem{Sommer:2014mea}
  R.~Sommer,
  PoS LATTICE {\bf 2013}, 015 (2014)
  doi:10.22323/1.187.0015
  [arXiv:1401.3270 [hep-lat]].





\bibitem{Soltz:2015ula} 
  R.~A.~Soltz, C.~DeTar, F.~Karsch, S.~Mukherjee and P.~Vranas,
  Ann.\ Rev.\ Nucl.\ 
  Part.\ Sci.\  {\bf 65}, 379 (2015)
  doi:10.1146/annurev-nucl-102014-022157
  [arXiv:1502.02296 [hep-lat]].







\bibitem{AliKhan:2001ek}
  A.~Ali Khan {\it et al.} [CP-PACS Collaboration],
  Phys.\ Rev.\ D {\bf 64}, 074510 (2001)
  doi:10.1103/PhysRevD.64.074510
  [hep-lat/0103028].

\bibitem{Ejiri:2009hq}
  S.~Ejiri {\it et al.} [WHOT-QCD Collaboration],
  Phys.\ Rev.\ D {\bf 82}, 014508 (2010)
  doi:10.1103/PhysRevD.82.014508
  [arXiv:0909.2121 [hep-lat]].


\bibitem{Bornyakov:2009qh}
  V.~G.~Bornyakov, R.~Horsley, S.~M.~Morozov, Y.~Nakamura, M.~I.~Polikarpov, P.~E.~L.~Rakow, G.~Schierholz and T.~Suzuki,
  Phys.\ Rev.\ D {\bf 82}, 014504 (2010)
  doi:10.1103/PhysRevD.82.014504
  [arXiv:0910.2392 [hep-lat]].
  
  
\bibitem{DeGrand:2018tzn}
T.~DeGrand, D.~C.~Hackett and E.~T.~Neil,
PoS \textbf{LATTICE2018}, 175 (2018)
doi:10.22323/1.334.0175
[arXiv:1809.00073 [hep-lat]].



\bibitem{Borsanyi:2012uq}
  S.~Borsanyi {\it et al.},
  JHEP {\bf 1208}, 126 (2012)
  doi:10.1007/JHEP08(2012)126
  [arXiv:1205.0440 [hep-lat]].

\bibitem{Borsanyi:2015waa}
  S.~Borsanyi {\it et al.},
  Phys.\ Rev.\ D {\bf 92}, no. 1, 014505 (2015)
  doi:10.1103/PhysRevD.92.014505
  [arXiv:1504.03676 [hep-lat]].

  
  
\bibitem{Giusti:1998wy}
L.~Giusti, F.~Rapuano, M.~Talevi and A.~Vladikas,
Nucl. Phys. B \textbf{538}, 249-277 (1999)
doi:10.1016/S0550-3213(98)00659-2
[arXiv:hep-lat/9807014 [hep-lat]].


\bibitem{Sharpe:1992ft}
S.~R.~Sharpe,
Phys. Rev. D \textbf{46}, 3146-3168 (1992)
doi:10.1103/PhysRevD.46.3146
[arXiv:hep-lat/9205020 [hep-lat]].


  
\bibitem{Blum:2001xb}
T.~Blum \textit{et al.} [RBC],
Phys. Rev. D \textbf{68}, 114506 (2003)
doi:10.1103/PhysRevD.68.114506
[arXiv:hep-lat/0110075 [hep-lat]].

\bibitem{Aoki:2005ga}
Y.~Aoki, T.~Blum, N.~H.~Christ, C.~Dawson, T.~Izubuchi, R.~D.~Mawhinney, J.~Noaki, S.~Ohta, K.~Orginos and A.~Soni, \textit{et al.}
Phys. Rev. D \textbf{73}, 094507 (2006)
doi:10.1103/PhysRevD.73.094507
[arXiv:hep-lat/0508011 [hep-lat]].


\bibitem{DeGrand:2007tx}
T.~DeGrand and S.~Schaefer,
[arXiv:0712.2914 [hep-lat]].

  

\bibitem{Lepage:1989hd}
  G.~P.~Lepage,
  ``The Analysis Of Algorithms For Lattice Field Theory,''
invited lectures at the 1989 TASI summer school, Boulder CO, June 4-30, 1989.
  CLNS-89-971.

\bibitem{Parisi:1983ae}
G.~Parisi,
Phys. Rept. \textbf{103}, 203-211 (1984)
doi:10.1016/0370-1573(84)90081-4


\bibitem{Clarke:2019tzf}
D.~A.~Clarke, O.~Kaczmarek, F.~Karsch and A.~Lahiri,
PoS \textbf{LATTICE2019}, 194 (2020)
doi:10.22323/1.363.0194
[arXiv:1911.07668 [hep-lat]].


\bibitem{Clarke:2020htu}
D.~A.~Clarke, O.~Kaczmarek, F.~Karsch, A.~Lahiri and M.~Sarkar,
Phys. Rev. D \textbf{103}, no.1, L011501 (2021)
doi:10.1103/PhysRevD.103.L011501
[arXiv:2008.11678 [hep-lat]].
 
 


\bibitem{Panero:2009tv} 
  M.~Panero,
  Phys.\ Rev.\ Lett.\  {\bf 103}, 232001 (2009)
  doi:10.1103/PhysRevLett.103.232001
  [arXiv:0907.3719 [hep-lat]].

\bibitem{MILC} {\tt https://github.com/milc-qcd/milc\_qcd/}

\end{thebibliography}
\end{document}